\documentclass[final]{agujournal2018}

\usepackage{amsfonts}
\usepackage{amssymb}
\usepackage{amsmath}
\usepackage{dsfont}
\usepackage[colorlinks,citecolor=red,urlcolor=blue]{hyperref}
\usepackage[caption=false,font=footnotesize]{subfig}

\usepackage{todonotes}

\usepackage{color}
\definecolor{red}{rgb}{1.0,0.,0.}

\usepackage[utf8x]{inputenc}
\graphicspath{{./figures/}}
\newcommand{\ignore}[1]{}

\newenvironment{mat}{\left[ \begin{array}{ccccccccccccc}}{\end{array}\right]}
\newenvironment{rmat}{\left[ \begin{array}{rrrrrrrrrrrrr}}{\end{array}\right]}
\newenvironment{lmat}{\left[ \begin{array}{lllllllllllll}}{\end{array}\right]}
\newcommand\bcm{\begin{mat}}
\newcommand\ecm{\end{mat}}
\newcommand\brm{\begin{rmat}}
\newcommand\erm{\end{rmat}}
\newcommand\blm{\begin{lmat}}
\newcommand\elm{\end{lmat}}

\newenvironment{pwdef}{\left\{ \begin{array}{ll}}{\end{array}\right.}
\newcommand\bpwdef{\begin{pwdef}}
\newcommand\epwdef{\end{pwdef}}

\newcommand\bc{\begin{center}}
\newcommand\ec{\end{center}}
\newcommand\bi{\begin{itemize}}
\newcommand\ei{\end{itemize}}
\newcommand\be{\begin{enumerate}}
\newcommand\ee{\end{enumerate}}

\newcommand\bsplit{\begin{split}}
\newcommand\esplit{\end{split}}

\newcommand{\eq}{\begin{equation}}
\newcommand{\en}{\end{equation}}
\newcommand{\eqm}{\begin{eqnarray}}
\newcommand{\enm}{\end{eqnarray}}
\newcommand{\eqmno}{\begin{eqnarray*}}
\newcommand{\enmno}{\end{eqnarray*}}
\newcommand{\eqml}[1]{\eql{#1}\begin{array}{rcl}}
\newcommand{\enml}{\end{array}\en}
\newcommand{\eql}{\begin{equation}\label}
\newcommand{\eqsub}[1]{\begin{subequations}\label{#1}\eqm }
\newcommand{\ensub}{\enm\end{subequations}}

\newcommand\reals{{{\rm l} \kern -.15em {\rm R} }}

\newcommand\qin{\Q_i^{n}}
\newcommand\qinp{\Q_i^{n+1}}

\newcommand\qijn{\Q_{ij}^n}
\newcommand\qijnp{\Q_{ij}^{n+1}}

\newcommand\apdq{{\mathcal A}^+\Delta \Q}
\newcommand\amdq{{\mathcal A}^-\Delta \Q}
\newcommand\bpdq{{\mathcal B}^+\Delta \Q}
\newcommand\bmdq{{\mathcal B}^-\Delta \Q}

\newcounter{equationgroup}
\newcommand\Q{Q}

\newcommand{\tFiphn}{\tilde F_{i+1/2}^n}
\newcommand{\tFimhn}{\tilde F_{i-1/2}^n}
\newcommand{\tFimhjn}{\tilde F_{i-1/2,j}^n}
\newcommand{\tFiphjn}{\tilde F_{i+1/2,j}^n}
\newcommand{\tGijphn}{\tilde G_{i,j+1/2}^n}
\newcommand{\tGijmhn}{\tilde G_{i,j-1/2}^n}

\newcommand{\apdqimh}{{\mathcal A}^+\Delta \Q_{i-1/2}}
\newcommand{\amdqiph}{{\mathcal A}^-\Delta \Q_{i+1/2}}

\newcommand\apdqimhj{\apdq_{i-1/2,j}}

\newcommand\amdqiphj{\amdq_{i+1/2,j}}

\newcommand\bpdqijmh{\bpdq_{i,j-1/2}}

\newcommand\bmdqijph{\bmdq_{i,j+1/2}}

\newcommand{\ico}[1]{#1 \kern -1ex \raisebox{1.1ex}{$\circ$}}
\newcommand{\icos}[2]{#1 \kern -1.1ex \raisebox{1.1ex}[.5em]{$\circ$}^{#2}}

\usepackage[]{apacite}
\usepackage{url} %this package should fix any errors with URLs in refs.
\draftfalse

\journalname{Journal of Advances in Modeling Earth Systems (JAMES)}

\begin{document}

%% ------------------------------------------------------------------------ %%
%  Title
%
% (A title should be specific, informative, and brief. Use
% abbreviations only if they are defined in the abstract. Titles that
% start with general keywords then specific terms are optimized in
% searches)
%
%% ------------------------------------------------------------------------ %%

\title{Efficient Tsunami Modeling on Adaptive Grids with Graphics Processing Units (GPUs)}

%% ------------------------------------------------------------------------ %%
%
%  AUTHORS AND AFFILIATIONS
%
%% ------------------------------------------------------------------------ %%

% Authors are individuals who have significantly contributed to the
% research and preparation of the article. Group authors are allowed, if
% each author in the group is separately identified in an appendix.)

% List authors by first name or initial followed by last name and
% separated by commas. Use \affil{} to number affiliations, and
% \thanks{} for author notes.
% Additional author notes should be indicated with \thanks{} (for
% example, for current addresses).

% Example: \authors{A. B. Author\affil{1}\thanks{Current address, Antartica}, B. C. Author\affil{2,3}, and D. E.
% Author\affil{3,4}\thanks{Also funded by Monsanto.}}
\authors{
    Xinsheng Qin\affil{1},
    Randall J. LeVeque\affil{2},
    Michael R. Motley\affil{1}
}

\affiliation{1}{Department of Civil and Environmental Engineering, University of Washington}
\affiliation{2}{Department of Applied Mathematics, University of Washington}

%% Corresponding Author:
% Corresponding author mailing address and e-mail address:

% (include name and email addresses of the corresponding author.  More
% than one corresponding author is allowed in this LaTeX file and for
% publication; but only one corresponding author is allowed in our
% editorial system.)
\correspondingauthor{Xinsheng Qin}{xsqin@uw.edu}

%% Keypoints, final entry on title page.

%  List up to three key points (at least one is required)
%  Key Points summarize the main points and conclusions of the article
%  Each must be 100 characters or less with no special characters or punctuation

\begin{keypoints}
\item A fast, accurate and GPU-based tsunami model on patched-based AMR is presented.
\item The model shows good relative performance with speed-ups of $3.6$ to $6.4$.
\item Absolute performance evaluation shows efficient usage of hardware resources.
\end{keypoints}

%% ------------------------------------------------------------------------ %%
%
%  ABSTRACT
%
% A good abstract will begin with a short description of the problem
% being addressed, briefly describe the new data or analyses, then
% briefly states the main conclusion(s) and how they are supported and
% uncertainties.
%% ------------------------------------------------------------------------ %%

\begin{abstract}
Solving the shallow water equations efficiently is critical to the study of natural hazards induced by tsunami and storm surge, since it provides more response time in an early warning system and allows more runs to be done for probabilistic assessment where thousands of runs may be required.  Using Adaptive Mesh Refinement (AMR) speeds up the process by greatly reducing computational demands, while accelerating the code using the Graphics Processing Unit (GPU) does so through using faster hardware.  Combining both, we present an efficient CUDA implementation of GeoClaw, an open source Godunov-type high-resolution finite volume numerical scheme on adaptive grids for shallow water system with varying topography.  The use of AMR and spherical coordinates allows modeling transoceanic tsunami simulation.  Numerical experiments on several realistic tsunami modeling problems illustrate the correctness and efficiency of the code, which implements a simplified dimensionally-split version of the algorithms.  This implementation is shown to be accurate and faster than the original when using CPUs alone.  The GPU implementation, when running on a single GPU, is observed to be $3.6$ to $6.4$ times faster than the original model running in parallel on a 16-core CPU. Three metrics are proposed to evaluate relative performance of the model, which shows efficient usage of hardware resources.
\end{abstract}

\section{Introduction}

Tsunamis are among the most dangerous natural disasters and have recently resulted in hundreds of thousands of fatalities and significant infrastructure damage (e.g. the 2004 Indian Ocean tsunami, the 2010 Chile tsunami, the 2011 Japan tsunami, and more recently the 2018 Indonesia tsunami).
A wide range of research has been conducted to simulate tsunami phenomena.
These studies are essential for evacuation planning \citep[e.g.,][]{scheer2012generic, lammel2010emergency}, 
designing vertical evacuation structures \citep[e.g.,][]{ash2015design,Gonzalez2013}, 
risk assessment \citep[e.g.,][]{gonzalez2013probabilistic,Adams2017,geist2006probabilistic,gonzalez2009probabilistic,annaka2007logic} 
and early warning systems \citep[e.g.,][]{taubenbock2009last, liu2009tsunami}.
The later two, in particular, can benefit from fast simulators. 
For instance, Probabilistic Tsunami Hazard Assessment (PTHA) often requires thousands of simulations to generate tsunami hazard curves and maps, and some early warning systems depend on rapid simulation results to allow longer evacuation time.

The numerical modeling of tsunamis often includes modeling multiple phases in very different scales, including tsunami generation from the source \citep[e.g.,][]{nosov2014tsunami,okada1985surface}, 
long-distance propagation \citep[e.g.,][]{George2006thesis,choi2003simulation,titov1997implementation}, 
local inundation of coastal regions \citep[e.g.,][]{park2013tsunami,qin2017multi,qin2018comparison}
and its interaction with coastal structures \citep[e.g.,][]{motley2015tsunami,qin2018three,winter2017tsunami}.
Some computer codes use separate models for different phases of tsunamis, while some integrate the modeling of multiple phases into a single simulation \citep[e.g.,][]{titov1997implementation,zhang2008efficient,macias2016comparison}, facing the computational challenges induced by very different scales (from thousands of kilometers to tens of meters) in the problem.

The simulation speed can be increased by either reducing computational cost, using more powerful machines, or both.
Since being proposed by \citet{berger1984adaptive}, the Adaptive Mesh Refinement (AMR) algorithm has been shown to effectively reduce computational cost in the numerical simulation of multi-scale problems. 
It can track features much smaller than the overall scale of the problem and adjust the computational grid during the simulation.
The algorithm has been implemented and developed into several frameworks and can be categorized into three major variants.
The first one is often referred to as patch-based or structured AMR \citep{Berger1989}. It allows rectangular grid patches of arbitrary size and any integer refinement ratios between two level of grid patches \citep[e.g.,][]{hornung2002samrai,bryan2014enzo,clawpack,zhang2016boxlib,adams2015chombo}.
Another variant is the cell-based AMR, which refines individual cells and
often uses a quadtree or octree data structure to store the grid patch information. 
The last variant is a combination of the first two, often referred to as block-based AMR. 
Unlike the patch-based AMR, which stores the multi-resolution grid hierarchy as overlapping and nested grid patches, this approach stores the grid hierarchy as non-overlapping fixed-size grid patches, each of which is stored as a leaf in a forest of quadtrees or octrees \citep[e.g.,][]{burstedde2014forestclaw,burstedde2011p4est,fryxell2000flash,macneice2000paramesh}.
In the past two decades, the AMR algorithm has been extensively applied for geophysical applications \citep[e.g.,][]{leng2011implementation,burstedde2013large,LeVeque2011}.
In particular, tsunami models that simulate both large scale transoceanic tsunami propagation and inundation of small-scale coastal regions save several orders of computational cost by using AMR.

Another approach to increasing tsunami simulation speed is to use faster hardware and/or a greater degree of parallelism to take advantage of modern architectures. 
Several codes parallelize tsunami modeling on multi-core CPUs \citep[e.g.,][]{pophet2011high}, and GeoClaw takes this approach via OpenMP. 
ForestClaw \citep{burstedde2014forestclaw}, on the other hand, can simulate a tsunami on distributed-memory machines with MPI parallelism.
Recently, use of the Graphics Processing Units (GPUs), which are even accessible on general desktop PCs, have become increasingly popular in the scientific computing community.
Many researchers have reported decent speed-ups in simulating tsunamis by using the GPUs.
However, most of these earlier studies involved implementing the PDE solvers on grids with constant spatial resolution, including those of \citet{acuna2009real}, \citet{Lastra2009}, \citet{de2011simulation}, \citet{brodtkorb2012efficient}, \citet{de2013efficient}, \citet{smith2013towards}, and \citet{DeLaAsuncion2016}.
This approach may suffer from too much computational cost when modeling transoceanic tsunami propagation, since no AMR is used.
The complexity of the AMR algorithm and data structure add challenges to the implementation of tsunami simulation model on the GPUs.
There have been some implementation of AMR algorithms on the GPUs with application to astrophysics \citep[e.g.,][]{wang2010adaptive,schive2010gamer}. 
However, very few models for simulating tsunami with AMR on the GPU have been developed. 
Relevant work includes simulation of landslide-generated tsunamis by \citet{de2017simulation} for example.

In this paper, a CUDA \citep{nickolls2008scalable} implementation of the patched-based AMR algorithm is developed and used to simulate tsunamis on the GPU. 
The Godunov-type wave-propagation scheme with 2nd-order limiters is implemented to solve the nonlinear shallow water system with varying topography.
Both canonical Cartesian grid coordinates for modeling tsunamis in small regions and spherical coordinates for transoceanic tsunami propagation are supported.
The use of AMR adds challenges to the implementation, including dynamic memory structure creation and manipulation, balanced distribution of computing loads between the CPU and the GPU, and optimizations to minimize global memory access and maximize arithmetic efficiency in the GPU kernel.
Numerical experiments on several realistic tsunami modeling problems are conducted to illustrate the correctness and efficiency of the implementation, showing speed-ups from 3.6 to 6.4 when compared to the original model running in parallel on 16-core CPU.

The paper is structured as follows. 
Section \ref{sec:swe} gives an overview of the shallow water equations and numerical schemes implemented to solve them.
Section \ref{sec:amr} briefly reviews the AMR algorithm and how it is combined with the numerical scheme described in Section \ref{sec:swe}. 
Section \ref{sec:implementation} describes the implementation details.
Section \ref{sec:results} shows the simulation results and performance statistics from several tsunami cases.
Finally the results are summarized in Section \ref{sec:conclusions}.

\section{Shallow Water Equation and Numerical Scheme}\label{sec:swe}
\subsection{Shallow Water Equation with Variable Topography}
The shallow water equations (SWEs) have been used broadly by many researchers in modeling of tsunamis, storm surge, and flooding.
It can be written in the form of a nonlinear system of hyperbolic conservation laws for water depth and momentum:

\begin{subequations}
\begin{align}
    \label{eq:swe_1}
    h_t + \left( hu \right)_x + \left( hv \right)_y & = 0, \\
    \label{eq:swe_2}
    \left( hu \right)_t + \left( hu^2 + \frac{1}{2}gh^2 \right)_x + \left( huv \right)_y &= -ghB_x - Dhu, \\
    \label{eq:swe_3}
    \left( hv \right)_t + \left( huv \right)_x + \left( hv^2 + \frac{1}{2}gh^2 \right)_y &= -ghB_y - Dhv,
\end{align}
\label{eq:swe}
\end{subequations}
where $u(x,y,t)$ and $v(x,y,t)$ are the depth-averaged velocities in the two horizontal directions, $B(x,y,t)$ is the bathymetry/topography,  and $D = D(h,u,v)$ is the drag coefficient.
The subscript $t$ represents a time derivative, while
the subscripts $x$ and $y$ represent spatial derivatives in the two horizontal directions.  The value of $B(x,y,t)$ is positive for topography above sea level and negative for bathymetry.
Coriolis terms can also be added to the momentum equations but is generally negligible for tsunami problems and is not used here.
The drag coefficient 
used in the current implementation is 
\begin{equation}
    D(h,u,v) = n^2gh^{-7/3}\sqrt{u^2+v^2},
\end{equation}
where $n$ is the \textit{Manning coefficient} and depends on the roughness of the ground.
A constant value of $n=0.025$ is often used for tsunami modeling, and this 
value is used for all benchmark problems in this study.

\subsection{Finite Volume Methods and Wave Propagation Algorithm}
% \subsubsection{Wave Propagation Algorithm}
A one-dimensional homogeneous system of hyperbolic equations in non-conservative form can be written as:
\begin{equation}
    q_t + A(q)q_x = 0,
    \label{eq:hyperbolic_non_conservative}
\end{equation}
For non-conservative form, the wave propagation algorithm \citep{LeVeque1997,rjl:fvmhp} can be used to update the solution:
\begin{equation}
    \qinp = \qin - \frac{\Delta t}{\Delta x} \left( \apdqimh + \amdqiph \right),
    \label{eq:wave_propagation_form}
\end{equation}
where $\apdqimh$ is the net effect of all right-going waves propagating into cell $\mathcal{C}_i$ from its left boundary, and $\amdqiph$ is the net effect of all left-going waves propagating into cell $\mathcal{C}_i$ from its right boundary.
Namely,
\begin{subequations}
\begin{align}
    \apdqimh & = \sum _{p=1}^m (\lambda ^p)^+ \mathcal{W}_{i-1/2}^p,\\
    \amdqiph & = \sum _{p=1}^m (\lambda ^p)^- \mathcal{W}_{i+1/2}^p,
\end{align}
\end{subequations}
where $m$ is total number of waves, $\mathcal{W}^p$ is the $p$th wave from the Riemann problem, $\lambda ^p$ is wave speed of the $p$th wave, and 
\begin{equation}
    (\lambda ^p)^+ = max(\lambda ^p,0), \quad (\lambda ^p)^- = min(\lambda ^p,0).
\end{equation}
The notations here are motivated by the linear case where $f(q) = Aq$. 
In such a case, the waves are simply decomposition of the initial jumps into basis form by the eigenvectors of the coefficient matrix $A$, propagating at the speed of eigenvalues:
\begin{equation}
    q_r-q_l = \sum _{p = 1}^m \mathcal{W}^p = \sum _{p = 1}^m \alpha ^pr^p,
\end{equation}
where $q_r$ and $q_l$ are right and left states of the Riemann problem, $r^p$ is the $p$th eigenvector of matrix $A$, and $\alpha ^p$ is coordinate in the direction of $r^p$.

% \subsubsection{Second-order Corrections and Wave Limiters}
The wave propagation form of Godunov's method (equation \ref{eq:wave_propagation_form}) is only first-order accurate and introduces a great amount of numerical diffusion into the solution.
This often smears out the steep gradients in the solution which are common in surface elevation near shoreline in tsunami simulation.
To obtain second-order resolution and maintain steep gradients, additional terms are added to equation \ref{eq:wave_propagation_form}:
\begin{align}
\begin{split}
    \qinp = \qin & - \frac{\Delta t}{\Delta x} \left( \apdqimh + \amdqiph \right) \\
                 & - \frac{\Delta t}{\Delta x} \left( \tFiphn - \tFimhn \right).
    \label{eq:wave_propagation_2nd_order}
\end{split}
\end{align}
The second-order correction terms are computed as
\begin{equation}
\tFimhn = \frac{1}{2}\sum _{p=1}^m \left( 1-\frac{\Delta t}{\Delta x}|\lambda ^p| \right) |\lambda ^p| \widetilde{\mathcal{W}}^p_{i-1/2},
\end{equation}
where the time step index $n$ is dropped and the superscript $p$ refers to the wave family.
The wave $\widetilde{\mathcal{W}}^p_{i-1/2} = \Phi(\theta ^p_{i-1/2})\mathcal{W}^p_{i-1/2}$ is a limited version of the original wave $\mathcal{W}^p_{i-1/2}$, where $\theta^p_{i-1/2}$ is a scalar that measures the strength of wave $\mathcal{W}^p_{i-1/2}$ relative to waves in the same wave family arising from a neighboring Riemann problem:
\begin{equation}
    \theta ^p_{i-1/2} = \frac{\mathcal{W}^p_{I-1/2} \cdot \mathcal{W}^p_{i-1/2}}{ \| \mathcal{W}^p_{i-1/2} \| },
    \label{eq:theta}
\end{equation}
where the index $I$ represents the interface on the upwind side of interface $x_{i-1/2}$
\begin{equation}
    I = 
    \begin{cases}
        i-1, & \text{if } \lambda^p_{i-1/2} > 0, \\
        i+1, & \text{if } \lambda^p_{i-1/2} < 0, 
    \end{cases}
\end{equation}
$\Phi(\theta)$ is a limiter function that gives values near 1 where solution is smooth and is close to 0 near discontinuities. 
Such property of a limiter function preserves second-order accuracy in region where the solution is smooth while avoiding non-physical oscillations arising near the discontinuities.
Note that computing limited waves adds complexity to the parallel algorithm implemented on the GPU since limited waves at one interface cannot be computed until its neighboring waves are solved. 
This is detailed in section \ref{sec:implementation}.

% \subsubsection{Algorithms in Two-Dimensional Space}
A two-dimensional hyperbolic system in non-conservative form
\begin{equation}
    q_t + A(q)q_x + B(q)q_y = 0
\end{equation}
is a general extension of the one-dimensional hyperbolic system (equation \ref{eq:hyperbolic_non_conservative}) in two-dimensional space.
The Godunov-type finite volume algorithms discussed above can be naturally extended to two-dimensional space by dimensional splitting, which splits the two-dimensional problem into a sequence of one-dimensional problems. 
The wave propagation algorithm now becomes
\begin{align}
\begin{split}
    Q_{ij}^{*} = \qijn & - \frac{\Delta t}{\Delta x} \left( \apdqimhj + \amdqiphj \right) \\
    & - \frac{\Delta t}{\Delta x} \left( \tFiphjn - \tFimhjn \right),
    \label{eq:wave_propagation_2nd_order_x}
\end{split}
\end{align}

\begin{align}
\begin{split}
    \qijnp = Q_{ij}^* & - \frac{\Delta t}{\Delta y} \left( \bpdqijmh + \bmdqijph \right) \\
    & - \frac{\Delta t}{\Delta y} \left( \tGijphn - \tGijmhn \right),
    \label{eq:wave_propagation_2nd_order_y}
\end{split}
\end{align}
where $\apdqimhj$ and $\amdqiphj$ are net effect of all waves propagating into cell $\mathcal{C}_{ij}$ from its left and right edges, 
while $\tFimhjn$ and $\tFiphjn$ are 2nd-order correction fluxes through its left an right edges. 
Similarly, 
$\bpdqijmh$ and $\bmdqijph$ are net effect of all waves propagating into cell $\mathcal{C}_{ij}$ from bottom and top edges,
while $\tGijmhn$ and $\tGijphn$ are 2nd-order correction fluxes through its bottom and top edges. 

In GeoClaw, the shallow water equation is written in non-conservative form, which augments the system by introducing equations for topography and momentum flux \citep{LeVeque2011}. 
An approximate Riemann solver has been implemented by \citet{George2008} to solve the Riemann problems at each cell interface for this augmented system.
This Riemann solver has some nice properties for tsunami modeling, including the capability of preserving steady state of the ocean, handling dry states in the Riemann problem, and maintaining non-negative depth in the solution.

% \subsubsection{Time Step Size}
The time step size $\Delta t$ for time integration must be chosen and adapted carefully at each time step if variable time step is used, which is typical for tsunami modeling.
The Courant, Friedrichs and Lewy (CFL) condition implies that the time step size for a certain AMR level must satisfy
\begin{equation}
    \nu \equiv \left| \frac{s \Delta t}{\Delta x} \right| \leq 1
\end{equation}
where $\nu$ is the CFL number and $s$ is the maximum wave speed seen at the AMR level.

\section{Adaptive Mesh Refinement}\label{sec:amr}
The block-structured adaptive mesh refinement algorithm implemented in Clawpack is
described in numerous papers, including \citet{berger1984adaptive} and \citet{Berger1989}, and is only briefly summarized here.

A collection of rectangular grid patches are used to store the solution.
Grid patches at different levels have different cell sizes.
The coarsest grid patches (level 1) cover the entire domain.
Grids patches at level $l$+1 are finer than coarser level $l$ grid patches by integer refinement ratios $r^l_x$ and $r^l_y$ in the two spatial directions, $\Delta x^{l+1} = \Delta x^l/r^l_x, \Delta y^{l+1} = \Delta y^l/r^l_y$, and cover sub-region of level $l$ grid patches.
In this study, the refinement ratios in the two spatial directions are always taken to be equal, $r^l_x = r^l_y$.
Typically, the time step size is also refined the same factor for level $l$+1 grid patches, $\Delta t^{l+1} = \Delta t^l/r^l_t$, with $r^l_t = r^l_x = r^l_y$.

The high level grid patches are regenerated every $K$ time steps such that they move with features in the solution.
When level $l$+1 grid patches need to be regenerated, some cells at level $l$ are flagged for refinement based on some criterion (in GeoClaw, typically where the amplitude of the wave is above some specified tolerance, or in specified regions where higher refinement is required, for example near the target coastal location, or where the wave will affect the solution in destination during time interval of interest as indicated by the backward adjoint solution\citep{davis2016adjoint}). 
The flagged cells are then clustered into new rectangular grid patches, which usually include some cells that are not flagged as well, using an algorithm proposed by \citet{Berger1991}.
The algorithm tries to keep a balance between minimizing the number of grid patches and minimizing the number of unflagged cells that are included in the resulting rectangular grid patches.
The newly generated level $l$+1 grid patches get their initial solution from either copying data from existing old level $l$+1 grid patches or, if no such grid patch exists, interpolating from level $l$ grid patches.
We say the level $l$+1 patch cells are ``on top'' of some level $l$ cells that cover the same spatial region.  
Note that the algorithm described below integrates the underlying level $l$ grid patches before the level $l+1$ patches.  After updating the finer patches, any level $l$ cells under level $l+1$ cells have their values updated to the average of level $l+1$ cell values.  After each regridding step, the new level $l+1$ patches need not cover the same level $l$ cells as previously.

\subsection{Time Integration} \label{sec:time_integration}
\begin{figure}[t]
\centering
\includegraphics[width=0.8\linewidth]{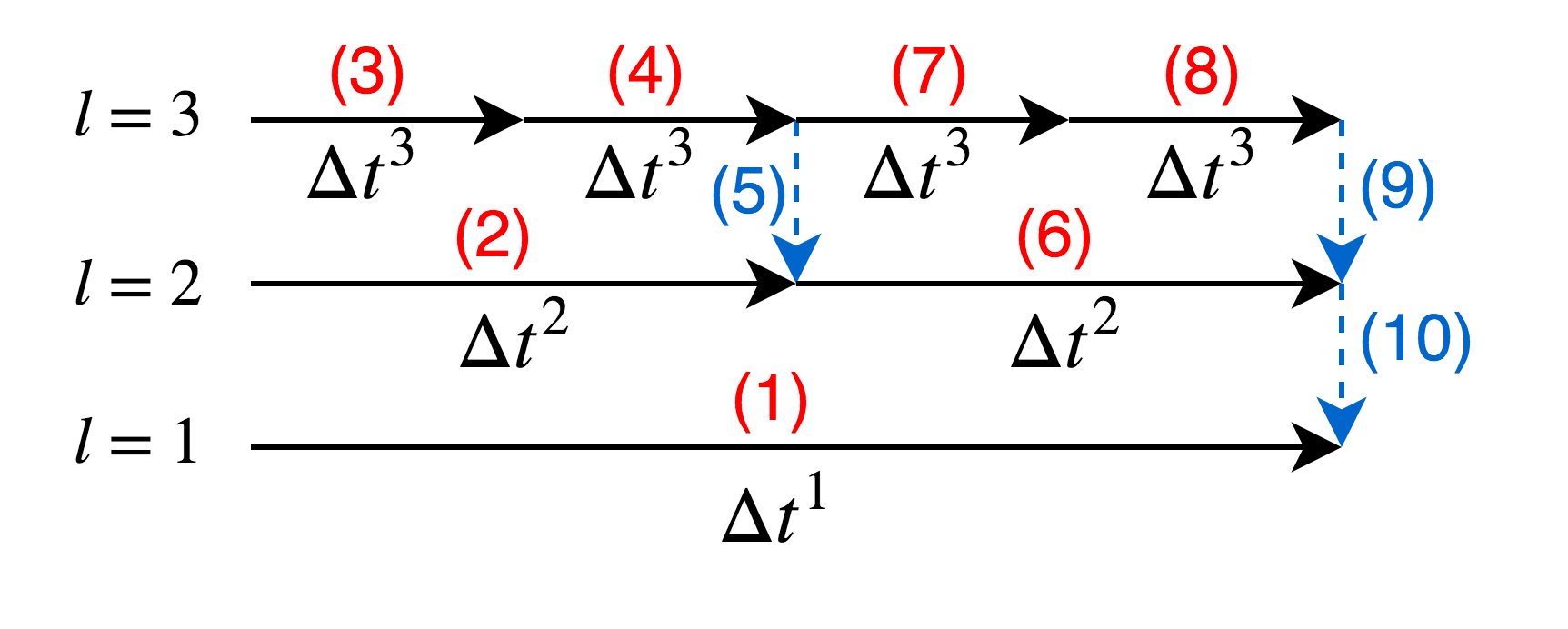}
\caption{Advancing the coarsest level by one time step, for a AMR hierarchy with 3 levels of grid patches. 
    The refinement ratio is 2 for both level $l=1$ and level $l=2$.
    Each black horizontal arrow in a solid line represents taking one time step on a specific level.
    Each blue vertical arrow in a dashed line represents one updating process that averages the solution from a fine level to a coarse level and refluxing process that preserves global conservation.
    The numbers from (1) to (10) describe the orders in which all operations are taken.
}
\label{fig:amr_time_integration}
\end{figure}

Each grid patch in the AMR grid hierarchy, despite different resolution, can be integrated in time with the wave-propagation form of Godunov's method described in the previous section.
Specifically, the following steps are applied recursively, starting from the coarsest grid patches at level $l=1$, as illustrated in a simple case in Figure \ref{fig:amr_time_integration}.
\begin{enumerate}
    \item Advance the solution in all level $l$ grid patches at $t_n$ by one step of length $\Delta t^l$ to get solution at $t_n+\Delta t^l$.
    \item Fill the ghost cells for all level $l+1$ grid patches, by either copying cell values from adjacent level $l+1$ grid patches if any exists, or interpolating in space and time from the cell values at level $l$ at $t_n$ and $t_n+\Delta t^l$ if no adjacent level $l+1$ grid patch exists.
Note that interpolation in time is generally required because the finer grids are integrated with smaller time steps.
    \item Advance the solution at level $l+1$ for $r^l_t$ time steps such that solution at level $l+1$ is at the same time as solution at level $l$. 
        Each time level $l+1$ is advanced, this entire algorithm (step 1--5) is applied recursively to the next finer level (with $l$ replaced by $l+1$ in these steps) if additional level(s) exist. 
    \item For any grid cell at level $l$ that is covered by level $l+1$ grid cells, the solution $Q$ in that cell is replaced with an appropriate weighted average of the values from the $r^l_x r^l_y$ level $l+1$ cells on top. This is referred to below as the {\em updating process}.
    \item For any grid cell at level $l$ that is adjacent to level $l+1$ grid cells, the solution $Q$ in that cell is adjusted to replace the value originally computed using fluxes found on level $l$ with the potentially more accurate value obtained by using level $l+1$ fluxes on the side of this cell adjacent to the level $l+1$ patch.  This step also preserves conservation for certain problems and is referred to below as the {\em refluxing process}. (This step is dropped in our implementation, see below.)
\end{enumerate}

Step 5 is important for some problems where exact conservation is expected, e.g., of a conserved tracer or for strong shock waves in nonlinear problems, and is necessary in this case to avoid the use of different numerical fluxes at the same interface on the side of the fine patch and the side where it abuts a coarser grid.  
However, this step requires storing additional flux information at every time step and communicating this information between levels and was found to have a large negative effect on the ability to speed up the code on the GPU.
We also found that this refluxing step has very little effect on the numerical results obtained for tsunami modeling (as shown in Section \ref{sec:japan2011}).  In this application we do not expect conservation of momentum at any rate (due to the topographic and friction source terms) and even conservation of mass is sacrificed when AMR is applied to a cell near the coast (as described in \citet{LeVeque2011}).  For these reasons we omit Step 5 in the GPU implementation.  
This greatly helps to optimize the logistics of the code and achieve very impressive performance in the benchmark problems, while only introducing negligible changes to the solution.

\subsection{Regridding}
Every time a level is advanced by $b$ time steps, regridding based on this level is conducted (except on the finest allowed level).
Typically, $b$ is chosen as 2--4 in tsunami modeling.
A larger $b$ results in less frequent regridding, which reduces time spent on the regridding process.
However, in order to ensure the waves in the solution do not propagate beyond the refined region before the next regridding process, when cells are flagged for refinement, usually an extra layer of $b$ cells surrounding the original flagged cells are flagged.
This makes each grid patch $2b$ cells wider in each of the two horizontal dimensions and thus introduces more cells, which increases the computational time of time integration.

In the regridding process, cells must be flagged before they are clustered into new grid patches.
A variety of different flagging criteria have been implemented, including flagging based on the slope of the sea surface, sea surface elevation, or adjoint methods.
For all the benchmarks in this study, the sea surface elevation is used for flagging.
In addition to this, some spatio-temporal regions might also be specified to enforce flagging in these regions, which is very useful for problems where both the transoceanic propagation and local inundation of a tsunami must be modeled, and thus require grid cells that are $O(10^3) \sim O(10^4)$ finer than the coarsest resolution in some near shore regions like a harbor or bay.

%\subsection{Interpolation and Averaging Strategies for Interpolation and Updating Process} \label{sec:interpolation_and_averaging}
During regridding, if a newly generated grid cell cannot copy values from old cells at the same level, its initial value must be interpolated from coarser levels.
In the updating process, coarse cell values get updated with the appropriate averaged value of fine grid cells on top.
An important requirement for both the interpolation and averaging strategies in tsunami modeling is to maintain the steady state of the ocean at rest, since refinement generally occurs before the tsunami waves arrive in an undisturbed area of the ocean.
For areas far from shoreline, the interpolation strategy can be simple linear interpolation and the averaging strategy used is averaging the surface elevation in fine cell values arithmetically and then compute depth in each cell based on the topography and surface elevation.
However, near the shoreline where one or more cells is dry, it is impossible to maintain conservation of mass and also preserve the flat sea surface during the interpolation or averaging in some circumstances. 
Details of the strategies can be found in \citet{LeVeque2011}.

\section{A Hybrid CPU/GPU Implementation}\label{sec:implementation}
One very basic question to answer in designing a GPU implementation of some code is which part of the program should be done by the CPU and which part should be done by the GPU.
In the current implementation we have put the Riemann solvers, wave limiter and CFL reduction on the GPU while letting the CPU take care of the rest, including the updating process, regridding process, filling ghost cells, and updating gauge values (finding the best grid patch to sample quantities of interest from, interpolating from cell values, and output), etc.
Since the GPU and the CPU considered in the study have separate physical memory, this design requires the transfer of solution data on each grid patch back and forth between the GPU and CPU memory through a PCI express 2.0 interface, which has relatively low bandwidth compared to the main memory of the CPU and the GPU. 
However, as we show later in the benchmark results, the extra time introduced by such data transfer takes less than $5\%$ of the total running time, since these operations can be carefully hidden by performing other operations concurrently.

If we instead put all procedures on the GPU, although it might save time by eliminating much of the data transfer, the code might suffer from having tiny GPU kernels that add significant overhead to the running time, and running inefficient GPU kernels that can be even slower than the CPU counterpart.
One example is filling ghost cells on the GPU. 
Each GPU kernel that fills ghost cells for a two-dimensional grid patch of $a$ by $a$ cells can have parallelism of $O(a)$ at most ($2a$ ghost cells to fill on each side), which is much less than the parallelism exposed in time integration of the same grid patch, which has parallelism of $O(a^2)$ and often involves much more computation.
The overhead of launching a GPU kernel is almost a fixed amount of time regardless of the actual execution time of the kernel. 
This overhead is often much longer than the execution time of a tiny GPU kernel like the GPU kernel that would be needed for filling ghost cells.
As a result, the total cost (including kernel launching overhead and kernel execution) of doing such operations on the GPU can be even higher than the cost of doing so on the CPU in some cases.
In addition to the consideration from kernel launching overhead, code that puts all procedures on the GPU wastes CPU computational resources.
Since a typical machine considered in this study consists of a multi-core CPU and a GPU, ideally the work load of the entire program would be distributed between the two as evenly as possible, such that the CPU stays busy as much as possible during the entire execution and so does the GPU.
In Section \ref{sec:results}, two metrics are proposed to measure such characteristics of a code in numerical experiments.

\subsection{Procedure Dependencies and Concurrent Execution}
% show dependency graph (this is like the pipeline)
% argue transferring data back and forth is okay
\begin{figure}[t]
\centering
\includegraphics[width=0.9\linewidth]{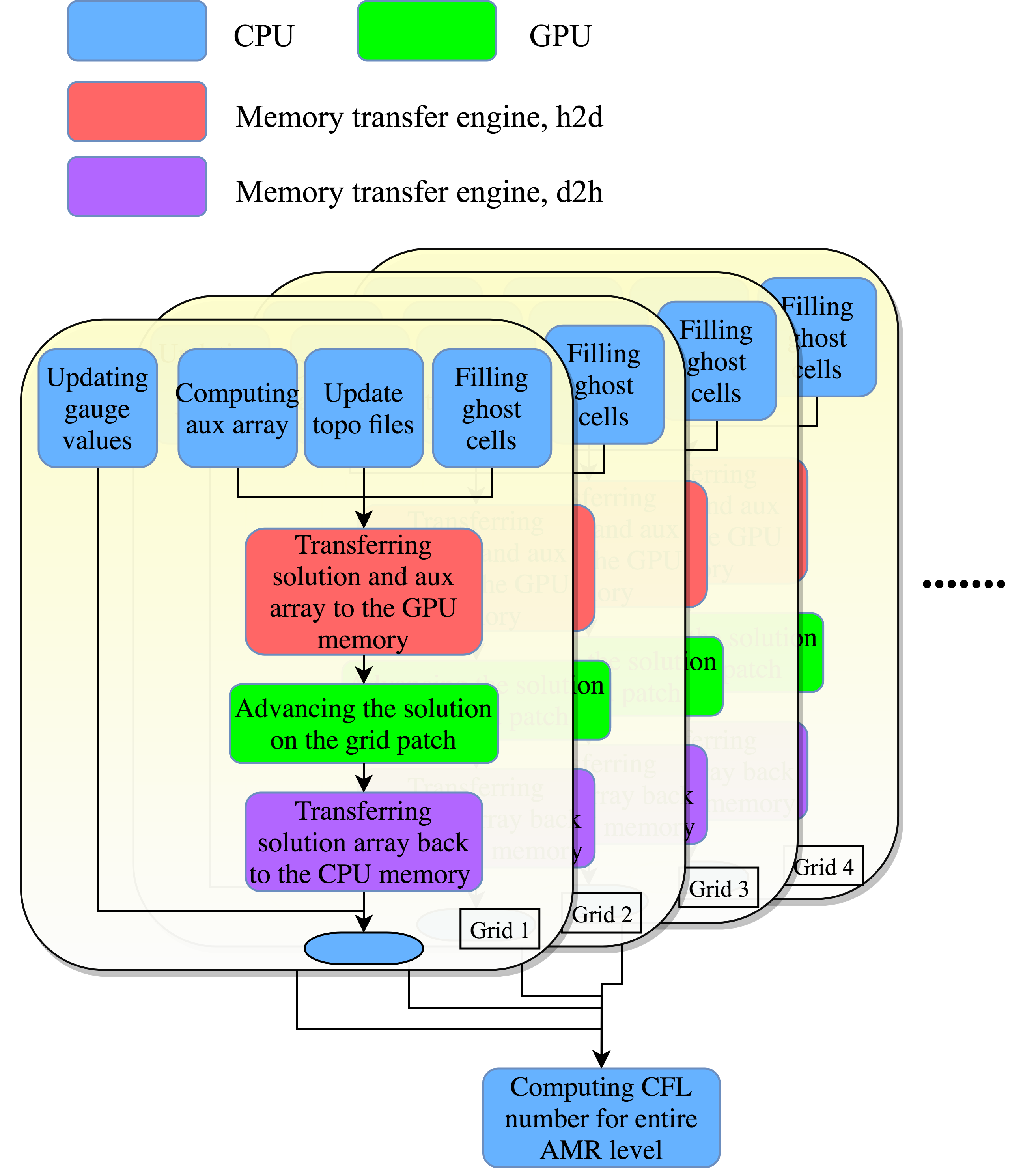}
\caption{Dependency graph of some major procedures in non-AMR portion of the code.
    The color indicates the hardware resource a procedure requires.
}
\label{fig:dependency_graph}
\end{figure}
Figure \ref{fig:dependency_graph} shows the major procedures of the code that are hearafter referred to as the {\em non-AMR portion} of the code.  (Omitted are the regridding process and updating process, essential components of AMR.)
An arrow from procedure $A$ to procedure $B$ indicates that procedure $A$ must be finished before procedure $B$ can start.
The color indicates the type of hardware resource a procedure needs.
4 colors represent 4 major types of hardware resource involved in the execution of the code. 
A blue block uses one CPU core, a green block uses the GPU Streaming Multiprocessors, a red block uses the memory transfer engine that transfers the data from the CPU memory to the GPU memory, and a purple block uses the memory transfer engine that transfers data in the opposite way. 
Note that these are separate transfer engines but there is only one of each.
Any two procedures without dependency can be done concurrently as long as relevant hardware resources are available.
The dependencies in the current implementation are enforced through a combination of rearrangement of CPU procedures and GPU kernel launches, use of OpenMP directives and CUDA streams, and proper synchronization between CPU threads and between the CPU and the GPU.

\begin{figure}[t]
\centering
\includegraphics[width=0.8\linewidth]{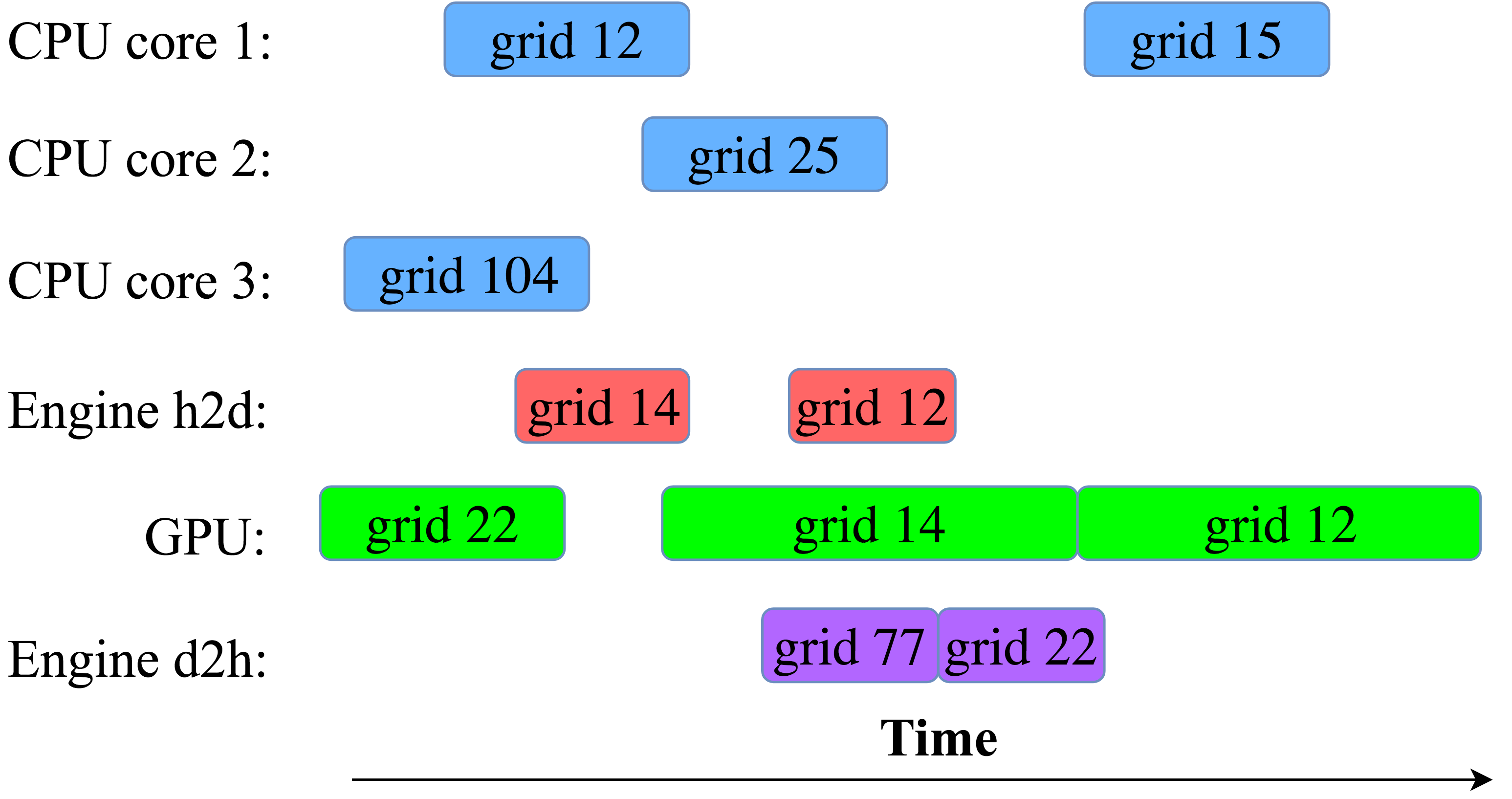}
\caption{
    An example of different procedures in non-AMR portion of the code running concurrently along the timeline. 
}
\label{fig:time_line}
\end{figure}

Figure \ref{fig:time_line} shows an example of these procedures being processed concurrently by four types of hardware on a machine with a three-core CPU.
Procedures follow the dependency specified in figure \ref{fig:dependency_graph}. 
Procedures that use the same hardware resource must wait in queue for the hardware to become available.
Note that procedures that needs CPU cores can use any available CPU core so those processed by different CPU cores can be executed concurrently. 

In figure \ref{fig:time_line}, during the entire time period when the red block for grid patch 12 is processed by one of the two memory transfer engines, the GPU Streaming Multiprocessors are processing the green block for grid patch 14.
In this case, transferring the solution data of grid patch 12 from the CPU memory to the GPU memory does not induce any extra cost.
During some middle time period of the red block for grid 14, however, no GPU computation is conducted so the GPU Streaming Multiprocessors are idle, which can be caused by unavailability of data on the GPU, for instance.
This time segment does induce extra cost due to transferring data between the CPU memory and the GPU memory.
Later in section \ref{sec:results}, such extra cost will be quantified to reveal the influence of transferring data between the CPU memory and the GPU memory on performance of the code.
Two additional metrics will also be defined and measured later in section \ref{sec:results}, the proportion of time during which the CPU has some work to do instead of waiting for the GPU to finish, and the proportion of time during which GPU Streaming Multiprocessors are doing computations.

\subsection{Memory Pool}
During regridding, new grid patches are generated and old ones are destroyed. 
New memory must be allocated on both the CPU and the GPU for storing solution data and auxiliary data for the new grid patches, while old memory for removed grid patches must be freed.
The total overhead of calling the CUDA runtime library to conduct these frequent memory operations cannot be neglected, and can even dominate execution of the code sometimes when grid patches are so small that the overhead is expensive relative to the time spent advancing the solution on grid patches.
To save the cost of such frequent memory operations, a memory pool is implemented, which requests a huge chunk of memory from the system by calling the CUDA runtime library at the initial time, keeping it until the end of execution, and getting more chunks when needed. 
All memory allocation and deallocation requests from the code are then through this memory pool at much less cost, with no need to actually allocate/free system memory.

% \subsection{Global Reduction for Adapting the Time Step Size}
% warp reduction
% Each time the solution on a grid patch is advanced on the GPU, a global reduction is conducted to collect the maximum wave speed seen on this grid patch.
% After the entire level is advanced, the maximum wave speed seen on each grid patch is further reduced to a single maximum value for the entire level on the GPU, which is then transferred back to the CPU memory for adjusting the time step size.

\subsection{Efficient Design of the Solver Kernels}

\subsubsection{The CUDA Programming Model}
The current implementation is based on the CUDA programming model and targeted Nvidia GPUs.
The architecture of Nvidia GPUs as well as explanation of the CUDA programming model are detailed in the Nvidia CUDA C programming guide \citep{nvidia}. 
Here only a brief review is given to provide sufficient knowledge for understanding the implementation details introduced in this section.

In the CUDA programming model, each function that is written to run on the GPU is called a CUDA kernel (or GPU kernel).
The code in a CUDA kernel specifies a set of instructions to be executed by multiple CUDA threads in parallel. 
The code can specify that some of the instructions should be executed by a certain group of threads but not the others.
All threads assigned to execute a CUDA kernel are grouped into CUDA blocks. 
All such CUDA blocks then form a CUDA grid.
CUDA blocks are independent of each other, can be sent to different Streaming Multiprocessors, and run concurrently.
During the execution of a CUDA kernel, each thread is provided with information regarding which CUDA block and which thread within that block it is. 
Based on this information, each thread can perform its own set of instructions on a specific portion of the data.

The GPU has many different types of hardware for data storage.
The three relevant types here are registers, shared memory and main memory.
The registers are the fastest storage and have very low access latency but each Streaming Multiprocessor has a very limited number of registers.
Each thread is assigned its own registers and thus can only access its own registers  (unless special instructions are used).

The shared memory has relatively slower bandwidth and longer access latency than the registers but its bandwidth is still much faster than that of the main memory and access latency is also much shorter than that of the main memory.
Each CUDA block is assigned a specified amount of shared memory, which is accessible by all CUDA threads in the CUDA block.
The quantity of registers and shared memory in the Streaming Multiprocessor is limited and fixed.
As a result, number of CUDA blocks that can reside in a Streaming Multiprocessor at the same time is limited by total number of registers and the amount of shared memory these CUDA blocks request.
If too few CUDA blocks can reside in a Streaming Multiprocessor at the same time, the Streaming Multiprocessor has low \textit{occupancy} and thus runs the CUDA kernel less efficiently.
For this reason, it is important to minimize the number of registers used by each thread and the amount of shared memory used by each CUDA block when the CUDA kernel is designed.

The main memory is located the farthest from the chip and thus has the lowest bandwidth and the longest access latency.
In principle, a CUDA kernel should have a minimal number of read and writes to main memory, especially given the fact that stencil computations for partial differential equations are often memory bandwidth bound.
A simple idea in CUDA kernel design is to load all data as efficiently as possible from the main memory to the shared memory, conduct all computations using the shared memory as a buffer to avoid unnecessary accesses of main memory, and then write new data back to the main memory.
However, this often causes too much shared memory usage for each CUDA block and results in very inefficient execution of the CUDA kernel.

% SOA vs AOS
\subsubsection{Data Layout}
Every 32 threads within a CUDA block are grouped as a warp, which executes the same instruction at the same time, including memory load and write operations.
The hardware can execute memory request from all threads in a warp most efficiently if they access a contiguous piece of memory.
This is called a coalesced access.
Such characteristic of the GPU hardware makes Structures-of-Arrays (SoA) preferable over Arrays-of-Structures (AoS). 
With SoA layout, the same state variables, e.g. water depth $h$, on the entire grid patch are stored contiguously in memory, 
whereas with AoS format, all state variables within the same grid cell are stored contiguously in memory.
Such a data layout results in strided access of the GPU memory. 
Namely, consecutive CUDA threads will access memory locations that are not consecutive. 
This can greatly reduce effective memory bandwidth since memory accesses cannot be coalesced. 

The current implementation contributes to the Clawpack eco-system \citep{Mandli2016}, which uses an AoS layout since Fortran arrays are dimensioned so that $q(m,i,j)$ is the $m$th component (depth or momenta) in the $(i,j)$ grid cell. 
However, many applications within the Clawpack eco-system will be affected if the AoS data layout is changed. 
Thus we continue to use the AoS layout in the current implementation. 
In the first half of the dimensional splitting method, the CUDA kernel that solves the equation in the $x$ direction reads in data in AoS but writes intermediate solution data in SoA, which is coalesced. 
The CUDA kernel that solves the equation in the $y$ direction then reads in data in SoA layout in a coalesced manner and writes new solution back in AoS layout.  

\subsubsection{CUDA Kernel Implementations}
% thread assignment
% - a bad idea: map thread to cells: redundant computation
% - map thread to cell interface. Launch a separate kernel to update
% - design used: combination of the two above, no redundant work, no separate kernel call
% - why not merge x-sweep and y-sweep: register overuse, lack of synchronization across CUDA blocks

In designing a CUDA kernel, one essential goal is to assign computational tasks to each thread.
To perform time integration on gird patches with Godunov-type wave-propagation methods, the goal is to decide how to distribute to each thread the tasks of solving the Riemann problems at each cell edge, limiting waves, and updating cell values.

\begin{figure}[t]
\centering
\includegraphics[width=0.9\linewidth]{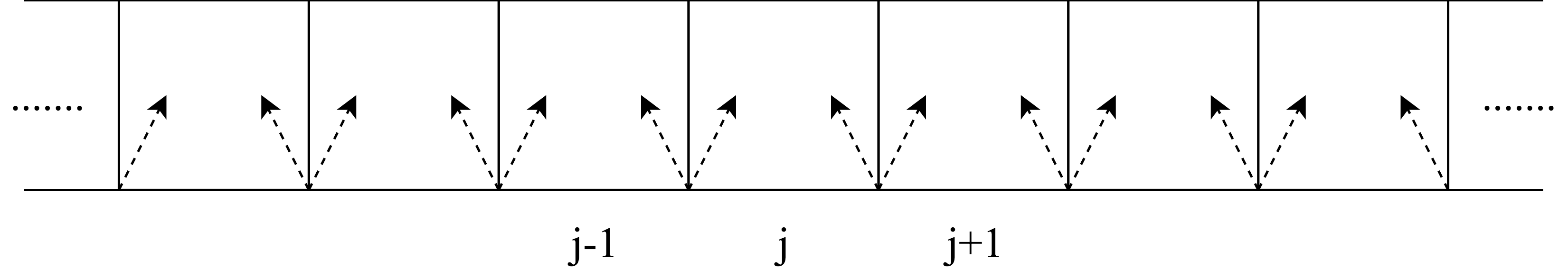}
\caption{
    A one-dimensional slice of a grid patch along the $x$ direction. 
    The arrows in dashed lines represent waves from the Riemann problems at the cell edges.
}
\label{fig:slice_1d}
\end{figure}

Figure \ref{fig:slice_1d} shows a one-dimensional slice of a grid patch along the $x$ direction when the first step of the dimensional splitting method is conducted to get the intermediate state $Q^*$.
Updating cell $C_{j}$ depends on the two sets of waves at cell edges $x_{j-1/2}$ and $x_{j+1/2}$.
When waves are limited, the waves at cell edge $x_{j-1/2}$ depends on waves at cell edges $x_{j-3/2}$ and $x_{j+1/2}$, while the waves at cell edge $x_{j+1/2}$ depends on waves at cell edges $x_{j+3/2}$ and $x_{j-1/2}$.
Any of these waves depends on the two cell values around them, respectively.
As a result, the solution at cell $C_{j}$ depends on four neighboring sets of waves, which depend on the 5 cell values around cell $C_j$ (including itself).

If each CUDA thread is assigned to update a cell in this one-dimensional slice, it needs to solve the four Riemann problems the cell depends on.
Redundant work is performed since some Riemann problems are solved and some waves are limited by neighboring CUDA threads as well.
On the other hand, if each thread is assigned to solve a Riemann problem at one cell edge in this one-dimensional slice, limit the waves, and then update the two neighboring cells with left- and right-going waves, the code must carefully avoid data racing since each cell is updated by two CUDA threads.
This typically involves the usage of a synchronization mechanism called lock, which decreases the execution efficiency of the CUDA kernel.

\begin{figure}[t]
\centering
\includegraphics[width=0.9\linewidth]{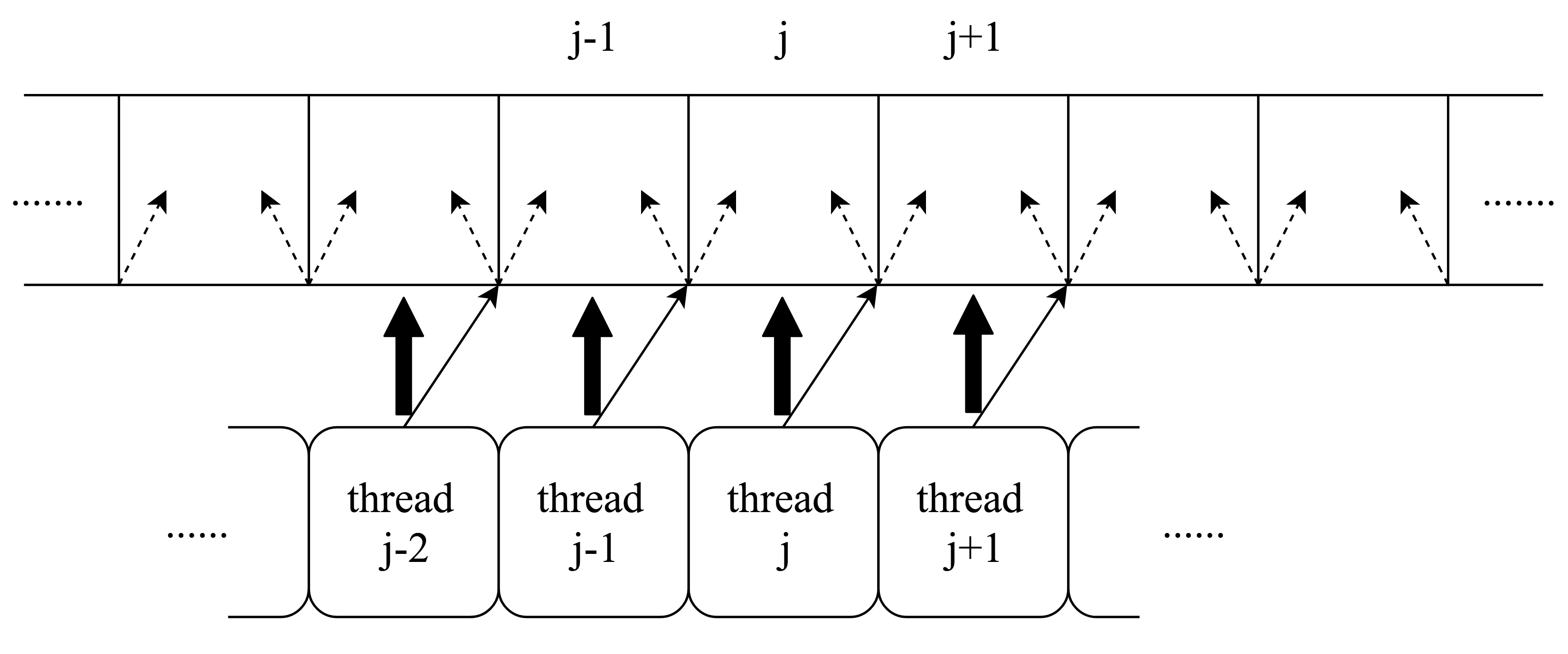}
\caption{
    Assign CUDA threads to grid cells and cell edges.
}
\label{fig:mapping}
\end{figure}

In the current implementation, a combination of the two ideas above is implemented.
Figure \ref{fig:mapping} shows how CUDA threads are first assigned to cell edges for solving Riemann problems and limiting waves, and then re-assigned to grid cells for updating the solution.
Each solid arrow denotes assigning one CUDA thread on a cell edge.
The thin arrows show the initial assignment to edges, while the thick arrows show the final assignment to cells.

In the first stage, each CUDA thread is assigned to a cell edge to solve the Riemann problem there. 
The left and right state for the Riemann problem for a thread is loaded from the main memory, while resulting waves from the Riemann problem are written into the shared memory.
In the second stage, each CUDA thread limits its waves to get correction fluxes at the cell edge it is assigned to. 
This requires reading waves from the two neighboring edges, which were produced by the two neighboring CUDA threads and stored in the shared memory.
Each thread then writes the limited waves (correction fluxes) back to the shared memory.

In the last stage, each CUDA thread is assigned to update a cell (thick arrows).
At this time, each thread already has left-going waves and correction fluxes at the right edge of its cell in its registers, which can be directly applied to update the cell value. 
The right-going waves and correction fluxes at the left edge of the same cell was produced by its left neighboring thread, and were stored in the shared memory in the last stage.
Thus each thread needs to read in these waves and fluxes from the shared memory and apply them to update the cell it is assigned to.
Each thread then writes the updated value $Q^*$ back to the main memory.
A similar kernel is then conducted for the second step of dimensional splitting method to get new state $Q^{n+1}$. 

This implementation requires a kernel to read solution data from the main memory only once at the beginning of the kernel execution and write the updated solution back to the main memory only once at the end of the kernel execution.
This is done by using only a reasonable number of GPU registers for each thread and a reasonable amount of shared memory for each CUDA block.
The usage of GPU registers for each CUDA thread only includes storing the state variables for left and right states of one Riemann problem, waves and wave speeds from one Riemann problem, plus any extra intermediate variables created during solving the Riemann problem,
while the usage of shared memory only includes waves from one Riemann problem per thread in a CUDA block.

\section{Numerical Results}\label{sec:results}

% \subsection{Machines Used for the Benchmarks}
This section is focused on evaluating the performance of current GPU implementation. 
Two machines for benchmarking the original CPU implementation and two machines for benchmarking the current GPU implementation are listed as below.
\begin{enumerate}
    \item a single Nvidia Kepler K20x GPU with a 16-core AMD Opteron 6274 CPU running at 2.2 GHz as the host;
    \item a single Nvidia TITAN X (Pascal) GPU with a 20-core Intel E5-2698 CPU running at 2.2 GHz as the host (but only 16 CPU threads are used for fair comparison with others);
    \item a single 16-core AMD Opteron 6274 CPU running at 2.2 GHz;
    \item a single 16-core Intel Xeon E-2650 CPU running at 2.0 GHz;
\end{enumerate}
As shown in the previous sections, the GPU implementation consists of jobs that are done by the CPU and jobs that are done by the GPU. 
In the benchmarks, the CPU implementation always runs in parallel with 16 OpenMP threads, using 16 CPU cores. 
The CPU part of the GPU implementation are also always processed in parallel by 16 OpenMP threads, using 16 CPU cores.
The GPU implementation solves the benchmark problems on machine 1 and 2 while the CPU implementation solves the same benchmark problem on machine 3 and 4.
Note that machine 1 and machine 3 have the same AMD CPU while machine 2 and machine 4 have similar Intel CPU.
Thus in this section, all speed-ups will be computed by comparing results on machine 1 to results on machine 3 and comparing results on machine 2 to results on machine 4.

We propose three metrics that can be used to measure absolute performance of current GPU implementation, which do not require comparing a GPU implementation to a CPU implementation.
We first define some quantities used in the definition of the three metrics.
Along the program execution time line $t\in\mathds{R}^+$, define $E^{GPU}_{i} = [t^{GPU}_{i,start}, t^{GPU}_{i,stop}]$ as the time interval that the $i$th GPU computation event (e.g. one of the green blocks in figure \ref{fig:time_line}) happens. 
$E^{GPU}_i$ is essentially a set of all moments that the $i$th GPU computation event is happening.
Similarly, define $E^{CPU}_{i}$, $E^{h2d}_{i}$ and $E^{d2h}_{i}$ for the $i$th CPU computation, the $i$th memory transfer from the CPU to the GPU memory and the $i$th memory transfer from the GPU to the CPU memory, respectively.
Then, all time intervals during which the GPU is doing computation, $\Omega ^{GPU}$, can be represented as $\Omega ^{GPU} = \bigcup_{i=1}^{N^{GPU}} E^{GPU}_i$, where $\bigcup$ is the union operation for sets and $N^{GPU}$ is total number of GPU computation events.
Similarly, we define another three sets of intervals for the other three types of events.
All time intervals during which the CPU is doing computation, $\Omega ^{CPU}$, can be represented as $\Omega ^{CPU} = \bigcup_{i=1}^{N^{CPU}} E^{CPU}_i$, where $N^{CPU}$ is total number of CPU computation events.
All time intervals  during which the memory transfer from the CPU memory to the GPU memory is happening, $\Omega ^{h2d}$, can be represented as $\Omega ^{h2d} = \bigcup_{i=1}^{N^{h2d}} E^{h2d}_i$, where $N^{h2d}$ is total number of such memory transfers.
All timer intervals during which the memory transfer from the GPU memory to the CPU memory is happening, $\Omega ^{d2h}$, can be represented as $\Omega ^{d2h} = \bigcup_{i=1}^{N^{d2h}} E^{d2h}_i$.where $N^{d2h}$ is total number of such memory transfers.
Lastly, define $E^{total} = [t_{start},t_{end}]$ as the time interval that the entire program runs.

The first metric measures the proportion of time during which the GPU is doing computation, defined as
 \begin{equation}
     P_1 = \frac{|\Omega ^{GPU}|}{|E^{total}|},
 \end{equation}
where $|\Omega|$ represents size of the set $\Omega$, which essentially computes the total length of all time intervals in $\Omega$ in this case.
Similarly, the second metric is defined as
 \begin{equation}
     P_2 = \frac{|\Omega ^{CPU}|}{|E^{total}|},
 \end{equation}
which measures proportion of time during which the CPU is doing computation. 
The last metric measures the proportion of extra time introduced by transferring data between the CPU memory and the GPU memory
 \begin{equation}
 P_3 = \frac{|\Omega ^{h2d} \bigcup \Omega ^{d2h} - \Omega ^{CPU} \bigcup \Omega^{GPU}|}{E^{|total}|},
 \end{equation}
where $-$ is the subtract operation for sets. 
For two sets $A$ and $B$, $A-B$ denotes all elements in $A$ but not in $B$.

\subsection{2011 Japan Tsunami}\label{sec:japan2011}
\subsubsection{Problem Setup}
The first benchmark problem is the 2011 Japan tsunami, which was triggered by an earthquake of magnitude 9.0-9.1 off the Pacific coast of Tōhoku, occurred at 14:46 JST (05:46 UTC) on Friday March 11th, 2011.
The earthquake source deformation files were obtained from NOAA Pacific Marine Environmental Laboratory (PMEL). 
They were not on an uniform latitude-longitude grid initially, and were converted to deformation information on uniform grids for use in our implementation.

\begin{figure*}[t]
\centering
\subfloat[]{\includegraphics[width=0.32\textwidth]{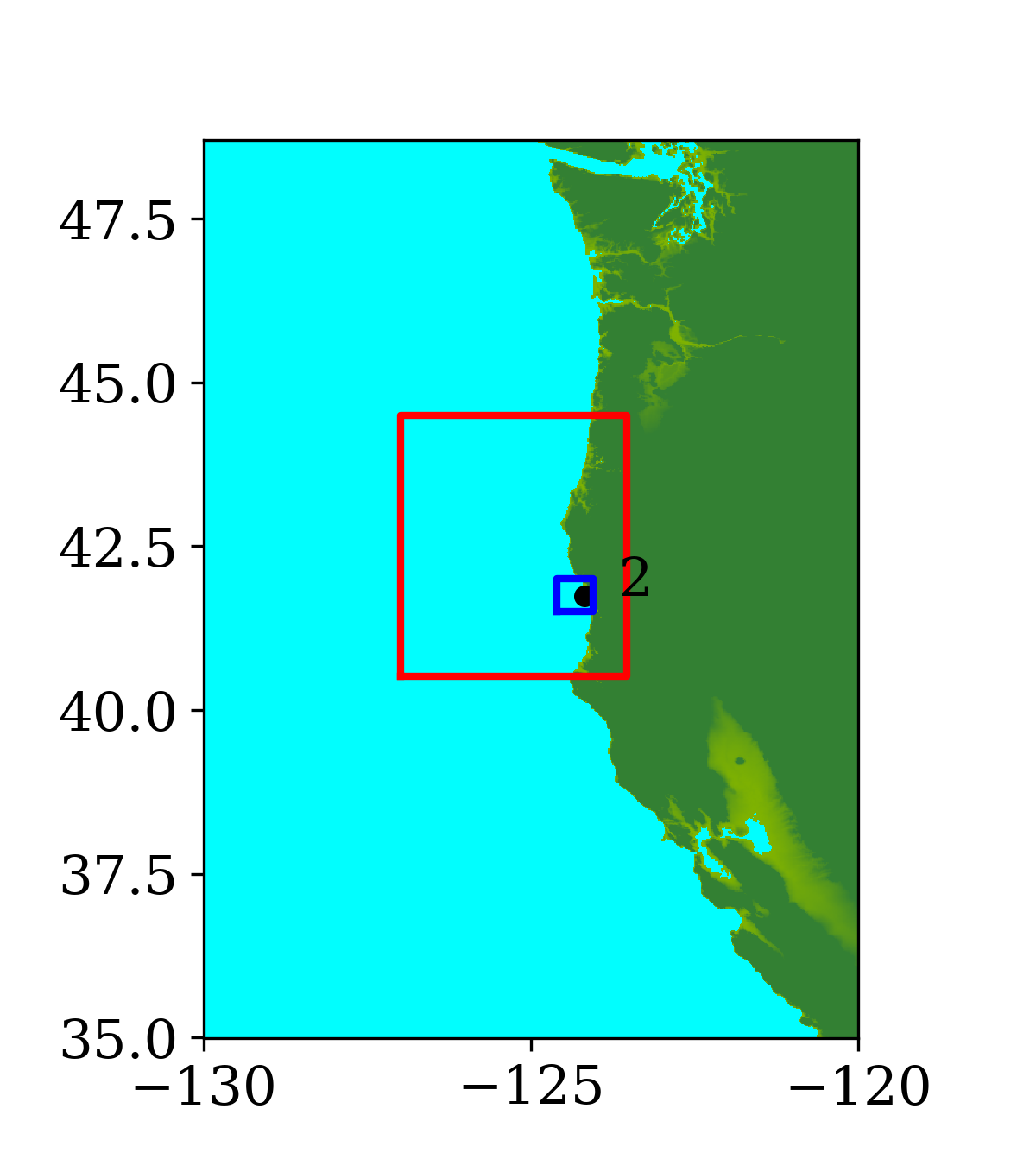} }
\hfill
\subfloat[]{\includegraphics[width=0.32\textwidth]{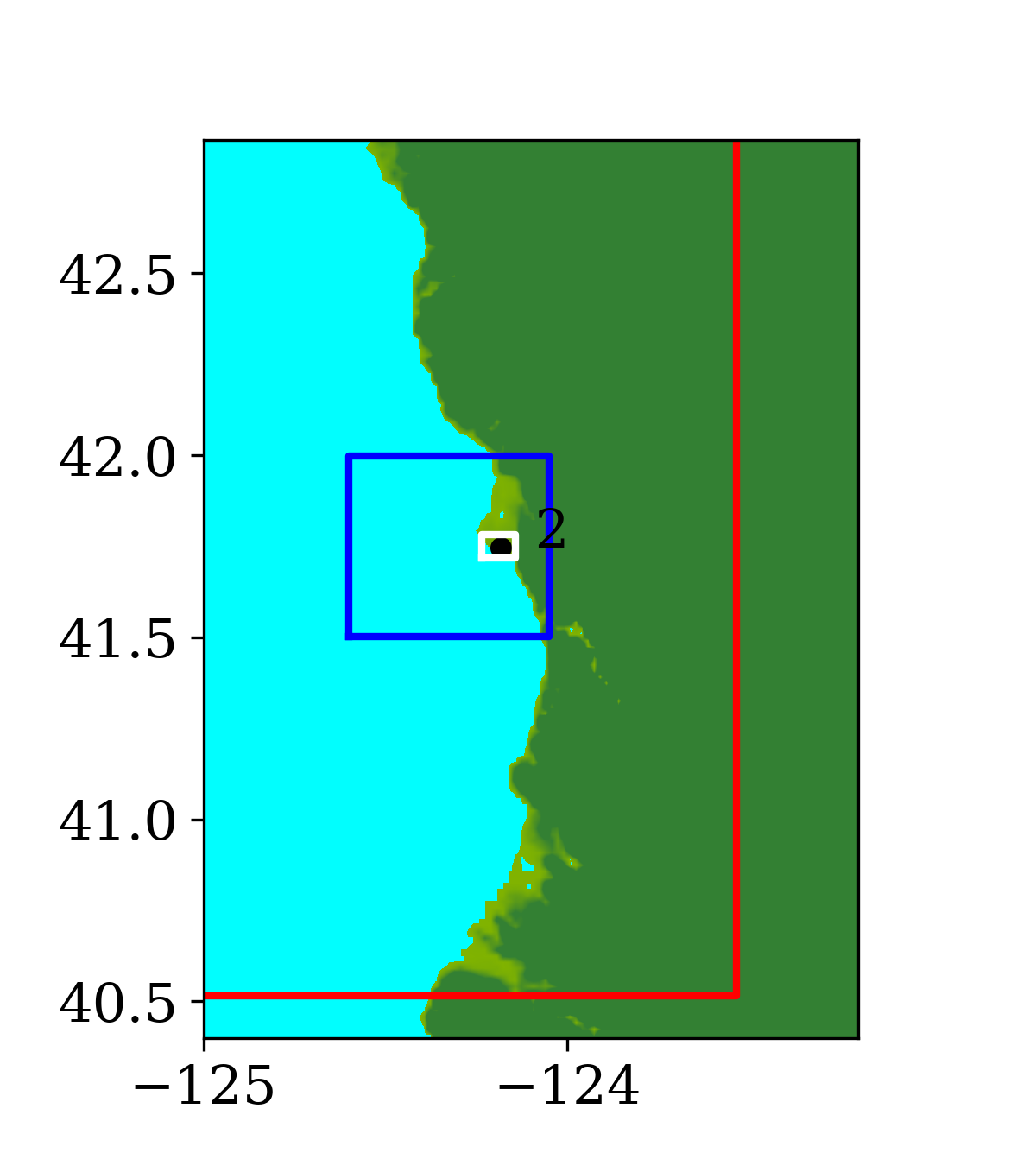} }
\hfill
\subfloat[]{\includegraphics[width=0.32\textwidth]{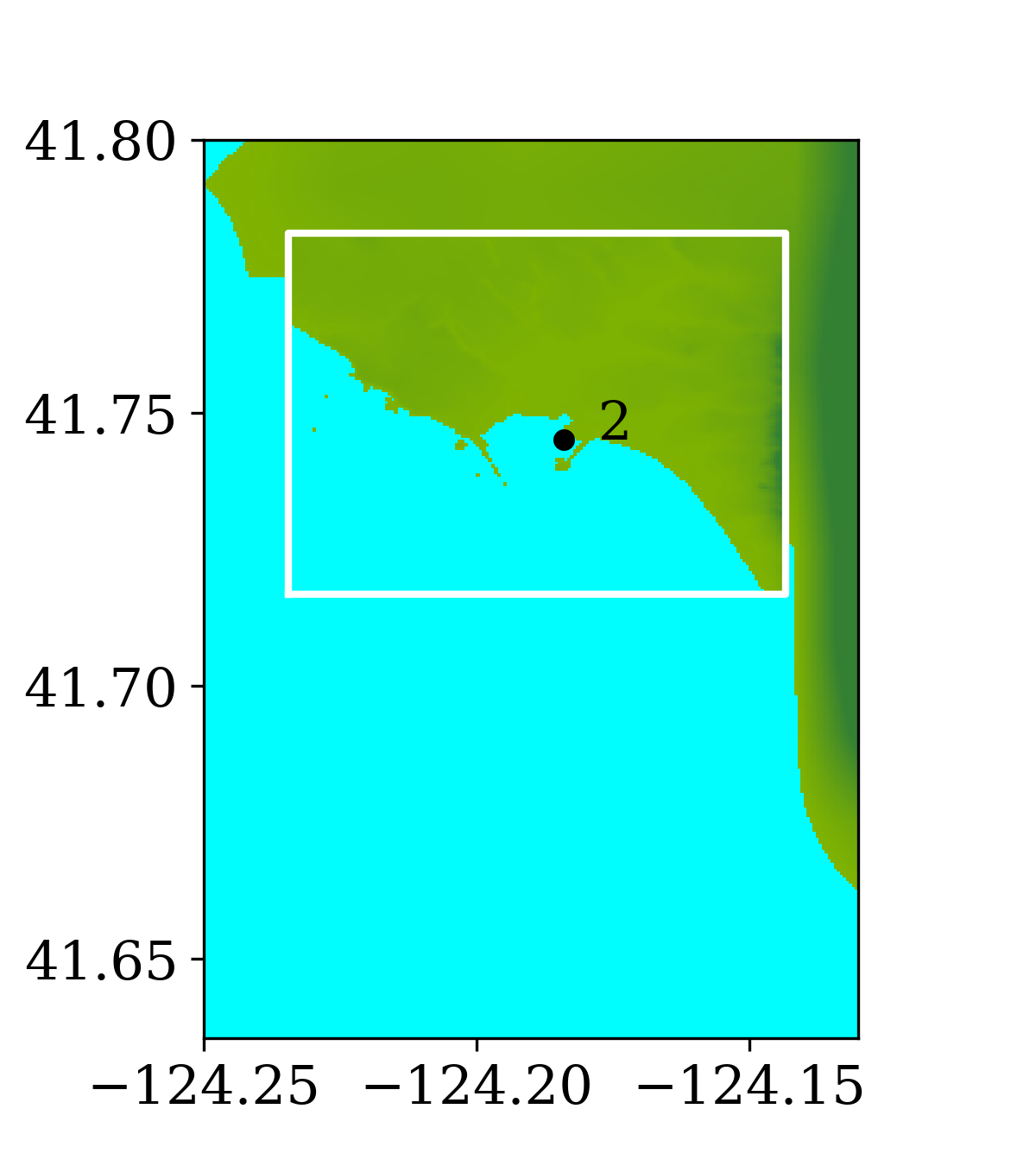} }
\caption{
    Three refinement regions around Crescent city with higher resolution and location of gauge 2. Red: level 4, 1-minute resolution; Blue: level 5, 12-second resolution; White: level 6, 2-second resolution.
     }
\label{fig:crescent_city}
\end{figure*}

The computational domain is from longitude $-240$ to $-100$ and latitude $-31$ to 65 in spherical coordinates.
Three levels of refinement are set across the ocean and around the source region (before getting close to the destination). 
Starting from the coarsest level (level 1) that has a resolution of 2 degrees, the refinement ratios are 5 and 6, giving a resolution of 25 minutes on level 2 and 4 minutes on level 3.
A refinement tolerance parameter can be specified to guide the mesh refinement. 
The smaller this parameter is set, the more likely the grid will be refined to the highest level allowed in a particular region. 
The refinement tolerance parameter is chosen to be the wave amplitude and is set to $0.005$ meter.
Thus, if a region is allowed to use any of the choices above (2 degree, 24 minute, 4 minute), the region will be refined up to a maximum of level 3 when the amplitude of a wave is higher than $0.005$.
In addition to specifying a tolerance for flagging individual cells, regions of the domain can be specified so that all cells in the region, over some time interval also specified, will be refined to at least some level and at most some level. 
In the simulation, the three refinement levels mentioned above are allowed in the entire region, with other constraints in specific sub-regions.
In the first 7 hours after the earthquake, a 4-minute resolution (level 3) in the region from longitude $-231$ to $-170$ and from latitude 18 to 62 is enforced.
This is reverted to the choices of 2 degrees or 24 minutes after 7 hours when the wave amplitudes are below tolerance.
Then moving onward toward the destination, a 4-minute resolution in the region from longitude $-170$ to $-120$ and from latitude 18 to 62 is enforced starting at 7 hours till the end of the 13-hour simulation.
Near Crescent City, the destination we are interested in, three higher higher levels of refinement regions are enforced to resolve for smaller-scale flow features near the coast:
\begin{enumerate}
    \item Level 4 with 1-minute resolution is enforced starting from 8 hours after the earthquake, in the region from longitude $-126.995$ to $-123.535$ and from latitude 40.515 to 44.495. 
    \item Level 5 with 12-second resolution is enforced starting from 8 hours after the earthquake, in the region from longitude $-124.6$ to $-124.05$ and from latitude 41.502 to 41.998.
    \item Level 6 with 2-second resolution is enforced starting from 8.5 hours after the earthquake, in the region from longitude $-124.234$ to $-124.143$ and from latitude 41.717 to 41.783.
\end{enumerate}
Figure \ref{fig:crescent_city} shows the three refinement regions as well as location of gauge 2, where the time series of water surface elevation is recorded. 

To ensure solution data on an entire grid patch can fit into the cache of the CPU for data locality, the size of each grid patch is limited to 128 by 128 for both the GPU and CPU cases.
The Godunov-type dimensional splitting scheme is used with 2nd order MC limiter applied to the waves.
The problem is simulated for a simulation time of $13$ hours.

\subsubsection{Simulation Results}
\begin{figure*}[t]
\centering
\subfloat[]{\includegraphics[width=0.49\textwidth]{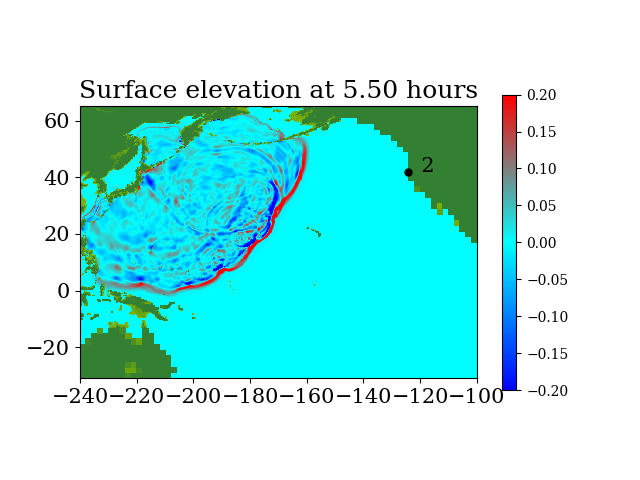} }
\hfill
\subfloat[]{\includegraphics[width=0.49\textwidth]{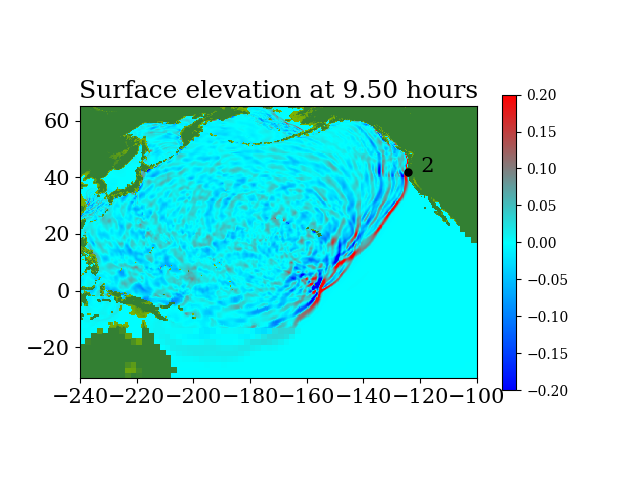} }
\caption{$\zeta(x,y,t)$ at $5.5$ hours and $9.5$ hours after the Japan 2011 earthquake.
     }
\label{fig:simulation_japan_2011}
\end{figure*}

\begin{figure*}[t]
\centering
\subfloat[]{\includegraphics[width=0.45\textwidth]{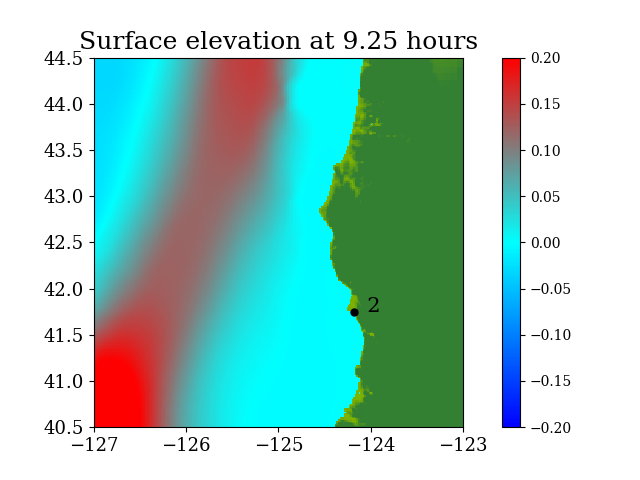} }
\hfill
\subfloat[]{\includegraphics[width=0.45\textwidth]{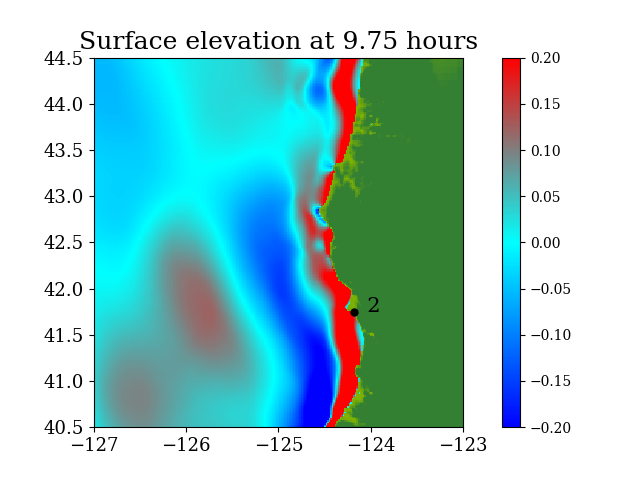} }
\caption{$\zeta(x,y,t)$ at $9.25$ hours and $9.75$ hours after the Japan 2011 earthquake, zoomed in near Crescent city.
     }
\label{fig:simulation_japan_2011_zoom}
\end{figure*}

Figure \ref{fig:simulation_japan_2011} and Figure \ref{fig:simulation_japan_2011_zoom} show snapshots during the simulation for the entire computational domain and near Crescent city, colored by $\zeta(x,y,t)$ defined as
\begin{equation}
    \zeta(x,y,t) = 
    \begin{cases}
        h(x,y,t), & \text{if } B(x,y) > 0 \text{  (the flow depth),} \\
        h(x,y,t)+B(x,y), & \text{if } B(x,y) \leq 0 \text{  ($\eta$, water surface elevation).}
    \end{cases}
    \label{eq:zeta}
\end{equation}

\begin{figure}[t]
\centering
\includegraphics[width=0.6\linewidth]{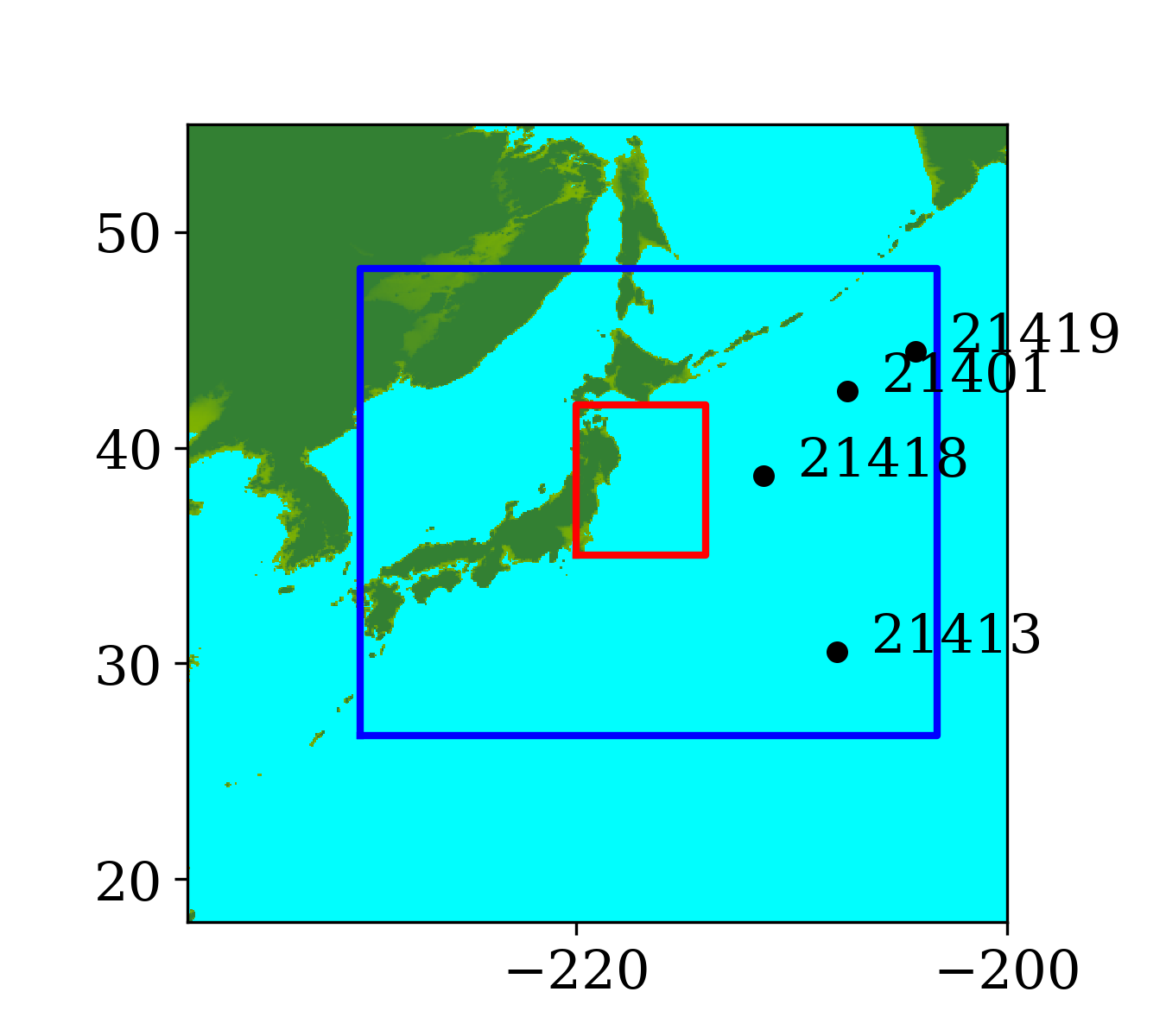}
\caption{Japan 2011 earthquake source and DART buoys locations. 
The coordinates for each DART buoys are:
1) gauge 21401, longitude $-207.417$, latitude $42.617$;
2) gauge 21413, longitude $-207.883$, latitude $30.515$;
3) gauge 21418, longitude $-211.306$, latitude $38.711$;
4) gauge 21419, longitude $-204.264$, latitude $44.455$.
}
\label{fig:dart}
\end{figure}

\begin{figure}[t]
\centering
\includegraphics[width=0.9\linewidth]{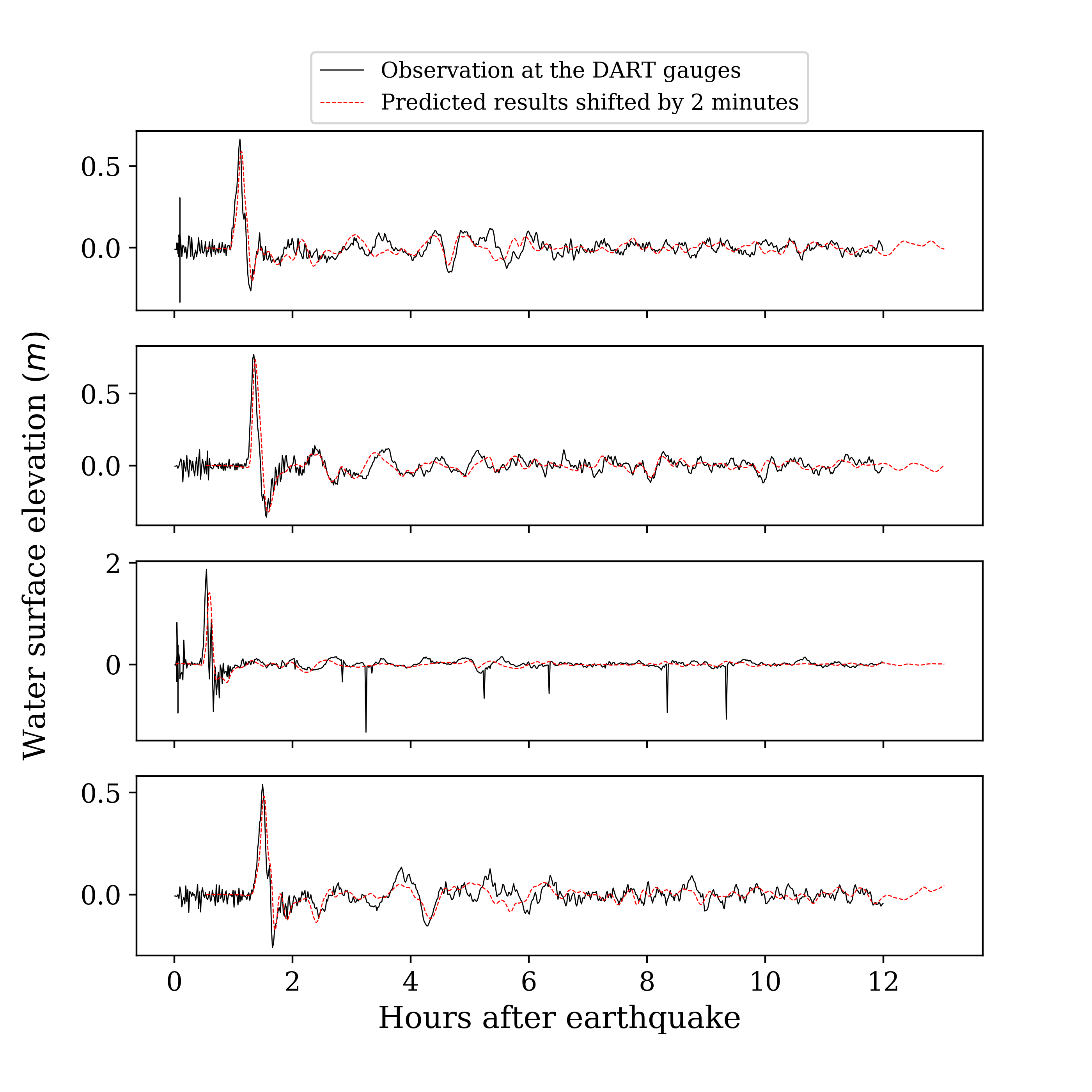}
\caption{Water surface elevation at 4 DART buoys. From top to bottom: gauge 21401, gauge 21413, gauge 21418, gauge 21419.}
\label{fig:wave_height_dart}
\end{figure}

During the tsunami, wave height data were recorded at 4 DART buoys (Deep-ocean Assessment and Reporting of Tsunamis) near the earthquake source, the locations of which are shown in Figure \ref{fig:dart}.
The blue rectangle in the figure indicates the extent of the earthquake source. 
However, most of the sea floor deformation is inside the red rectangular region, where one-minute topography files are used to make sure the region is well resolved.
The wave heights predicted by the current GPU implementation at the 4 DART buoys are shown in figure \ref{fig:wave_height_dart} and compared against observed data.
The comparison shows the predicted results agree quite well with observed data at the 4 DART buoys.

\begin{figure}[t]
\centering
\includegraphics[width=0.95\linewidth]{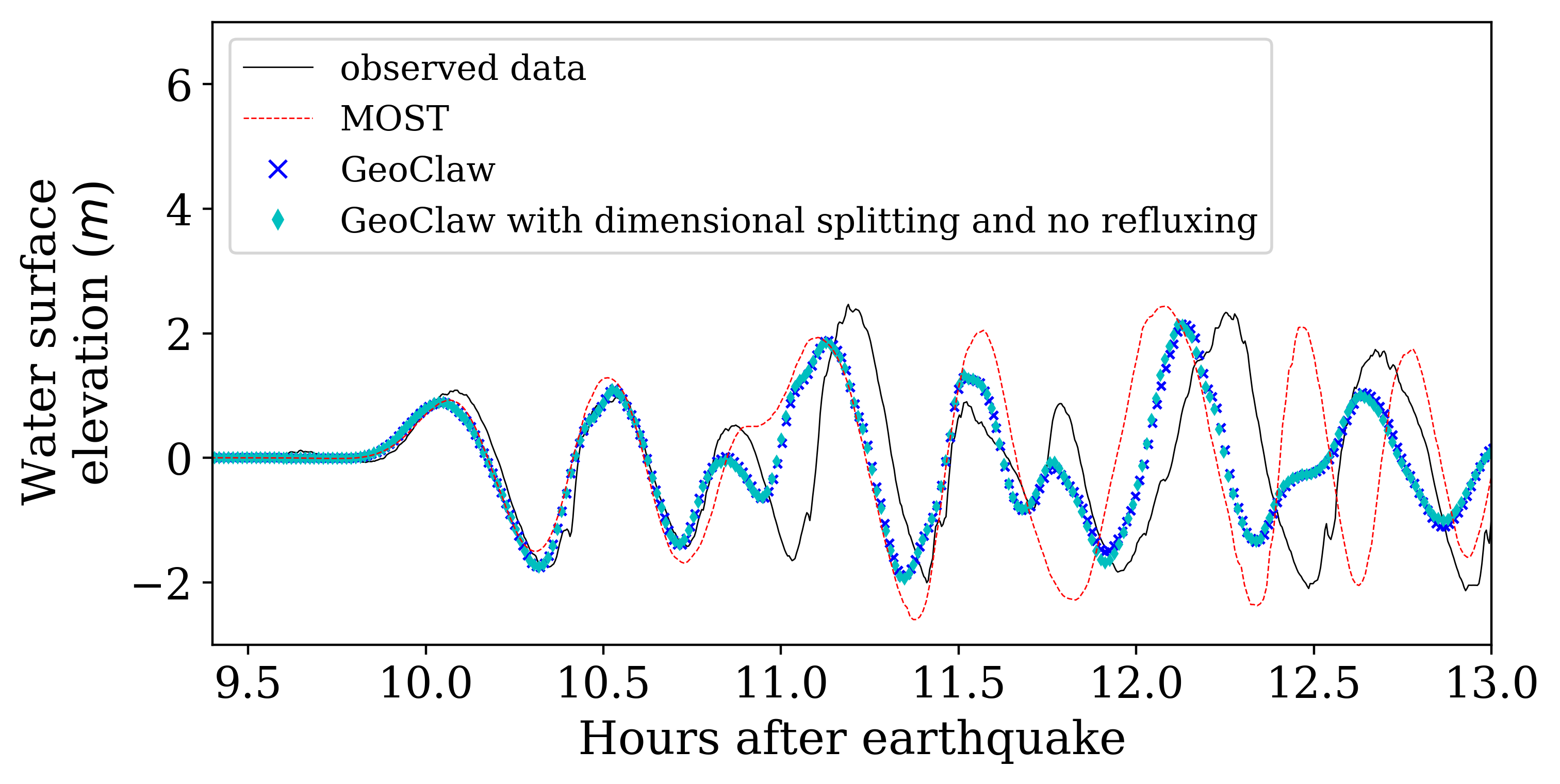}
\caption{Water surface elevation at gauge 2 (location: longitude $-124.1840$, latitude $41.7451$) near Crescent city. 
Time series from the MOST model are shifted by 6 minutes.
All other time series from numerical results are shifted by 6.5 minutes.}
\label{fig:wave_height_crescent_gauge}
\end{figure}

Recall that the current implementation uses a dimensional splitting scheme with no refluxing. 
To show that this simplification gives comparable results to the original GeoClaw code, Figure \ref{fig:wave_height_crescent_gauge} gives time series of surface elevation recorded at a tide gauge near Crescent City, California, United States, during the 2011 Japan Tsunami.
The observation has been detided by subtracting the predicted tide level from it to remove the influence of tide level.
Sample result from another well-known tsunami model MOST \citep{titov1997implementation} have also been included for comparison.
The comparison shows that a simplified GeoClaw CPU code that implements such a dimensional splitting scheme with no refluxing gives very close results to those produced by the original GeoClaw and agree well with another model and observed data.
The current GPU implementation gives identical results to this simplified version of the original GeoClaw and is not shown in the Figure.

\begin{figure}[t]
\centering
\includegraphics[width=1.0\linewidth]{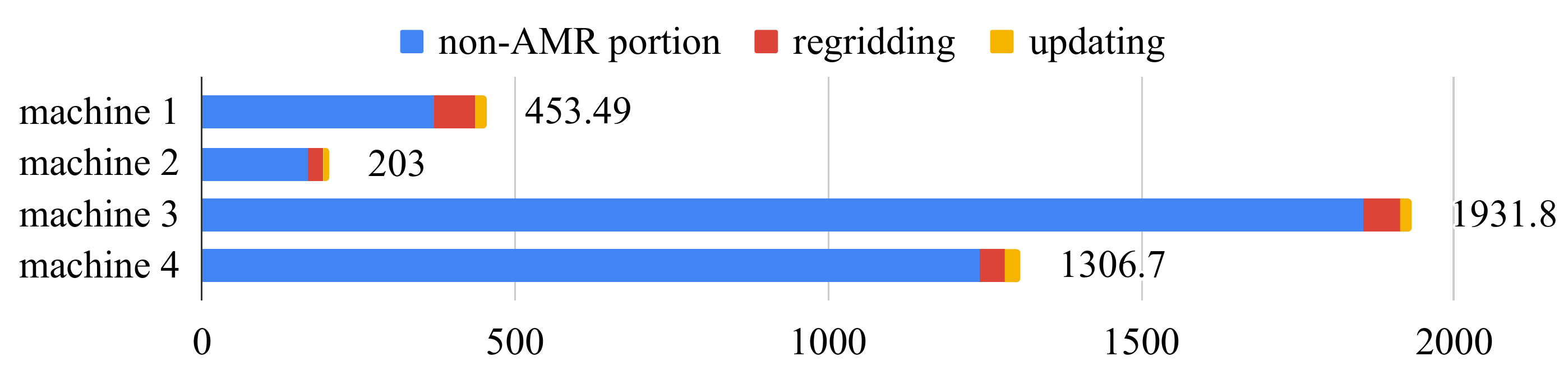}
\caption{Wall time (in seconds) of entire program on simulating the Japan 2011 tsunami, for original CPU implementation running on machine 3 and machine 4, and current GPU implementation running on machine 1 and machine 2. 
}
\label{fig:japan_total_time}
\end{figure}

% \subsubsection{Relative Performance}
Figure \ref{fig:japan_total_time} shows total running time and proportion of the three components on 4 machines.
The total speed-ups are $4.3$ on machine 1 and $6.4$ on machine 2 for current GPU implementation.
Note that since time spent on the non-AMR portion decreases on machine 1 and 2, the cost for regridding and updating take up larger portion of the total run time.
However, one could still only gain a very limited additional performance increase if the regridding and updating processes were implemented on the GPU.
Amdahl's law states theoretical speed-up of the execution of a whole program is
\begin{equation}
    S(s) = \frac{1}{(1-p)+\frac{p}{s}},
    \label{eq:amdahl}
\end{equation}
where $S$ is theoretical speed-up of the execution of the whole program, $s$ is the speed-up of the portion that is accelerated, $p$ is proportion of total running time that the accelerated portion takes.
Further more, one has $S(s) \leq \frac{1}{1-p}$,
where the equality is achieved when $s$ approaches $\infty$ in equation \ref{eq:amdahl}.
From Amdahl's law, even if the regridding and updating processes are implemented on the GPU and are accelerated infinitely, the entire program only get roughly $1.2$ speed-up on machine 1 and 2.
% The speed-ups for the entire non-AMR portion are $5.0$ for machine 1 and $7.3$ for machine 2.

Table \ref{tb:metrics_japan} shows the three metrics for current GPU implementation running on machine 1 and machine 2 when the Japan 2011 tsunami is simulated.
The proportion of GPU computation ($P_1$) reaches about $50\%$ on machine 1 and a higher percentage of $64\%$ on machine 2.
This could be due to the fact that machine 2 has a newer GPU which has much lower overhead for kernel launch and memory transfer.
The proportion of CPU computation ($P_2$) are around $80\%$ for both machines.
In other words, during $20\%$ of the total running time, the CPU is idle. 
$P_3$, the extra time introduced by transferring data between the CPU and the GPU memory, is less than $5\%$ for both machines. 
This shows that even if the data can be transferred infinitely fast between the CPU and the GPU memory so the data transfer has no effect on execution time at all, the total running time of the entire program can be reduced by at most $5\%$.
Thus having to transfer data between the CPU and the GPU memory is not a critical issue that affects the performance of the code.

% \subsubsection{Absolute Performance}
\begin{table}[t]
\caption{
    The three metrics measured from simulating the Japan 2011 tsunami on machine 1 and machine 2.
}
\centering
\begin{tabular}{|l|l|l|}
\hline
   & machine 1 & machine 2 \\ \hline
$P_1$ & 46.92\%   & 64.20\%   \\ \hline
$P_2$ & 84.50\%   & 79.30\%   \\ \hline
$P_3$ & 3.98\%    & 2.90\%    \\ \hline
\end{tabular}
\label{tb:metrics_japan}
\end{table}

%\clearpage % create all the figures before starting next section

\subsection{A local Tsunami Triggered by Near-field Sources}
\subsubsection{Problem Setup}
\begin{figure}[t]
\centering
\includegraphics[width=0.7\linewidth]{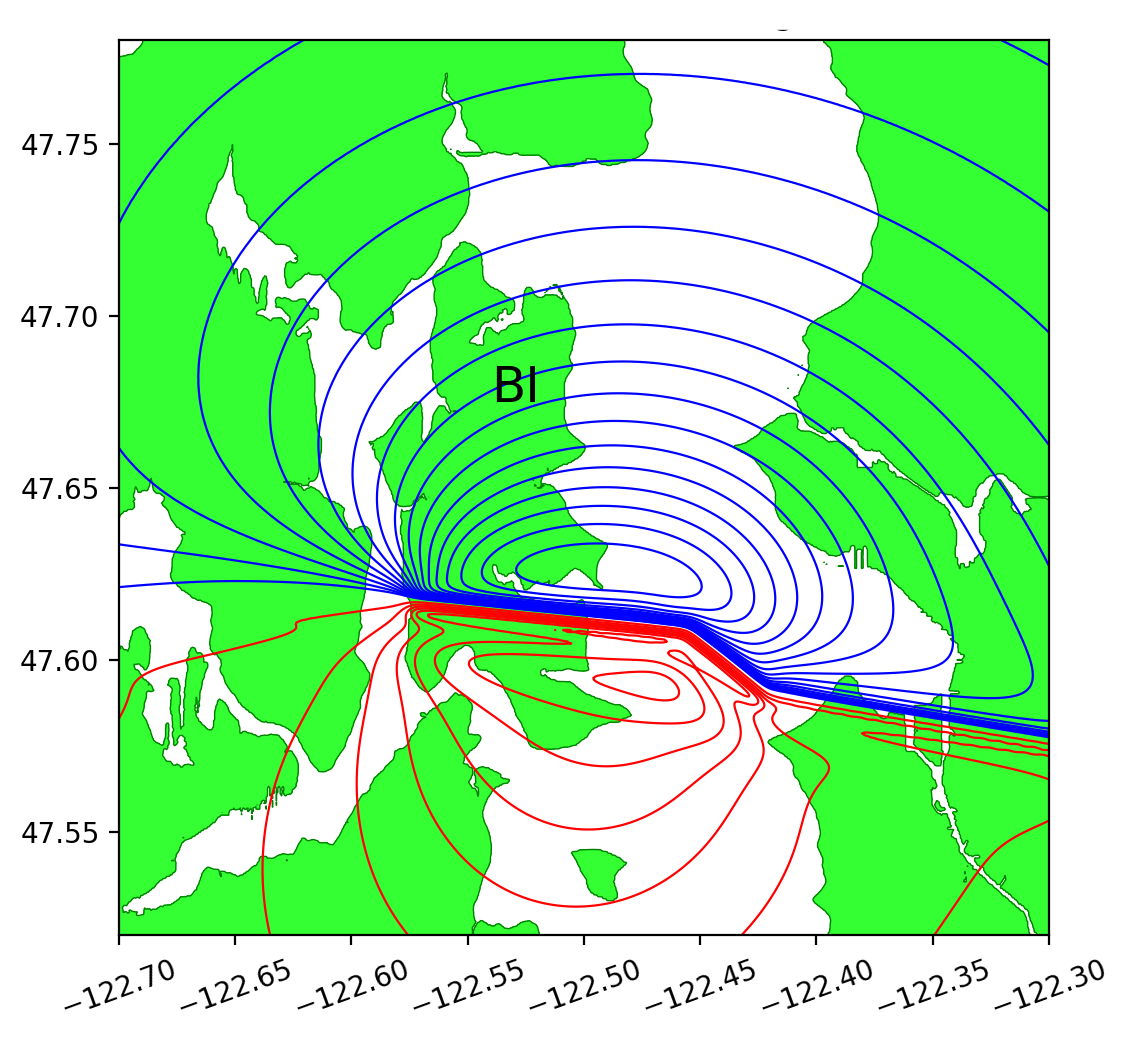}
\caption{
    Surface displacement for the hypothetical Seattle Fault earthquake, with Bainbridge Island labelled BI. Eagle Harbor is just north of the fault on the east side of the island.  
    Red contours show uplift at levels $0.5,~1,~1.5,~\ldots$ meters,
    blue contours show subsidence at levels
    $-0.05,$ $-0.1,~\ldots$ meters.
}
\label{fig:seattle_fault}
\end{figure}

The second benchmark problem is the modeling of a local tsunami that is triggered by a near-field earthquake, which typically hits the shoreline much earlier than a tsunami triggered by a far-field earthquake.
The tsunami is triggered by a hypothetical Mw 7.3 earthquake on the Seattle Fault, which cuts across Puget Sound 
(through Seattle and Bainbridge Island, see figure \ref{fig:seattle_fault}) 
and can create a tsunami that can cause significant inundation and high currents in some coastal communities around the Puget Sound.
The event was designed to model an earthquake that occurred roughly 1100 years ago, and for which geologic data is available for the uplift or subsidence at several locations.  
Here, we focused on modeling this local tsunami and predicting its impact on Eagle Harbor at the Bainbridge island, the location of which is shown below in figure \ref{fig:eagle_domain}.
The ground deformation file for generating the tsunami is obtained from PMEL, which has been used for recent comparison study of GeoClaw and MOST as part of a tsunami hazard assessment of Bainbridge Island \citep{THA_Bainbridge2018}.

\begin{figure*}[t]
\centering
\subfloat[]{\includegraphics[width=0.32\textwidth]{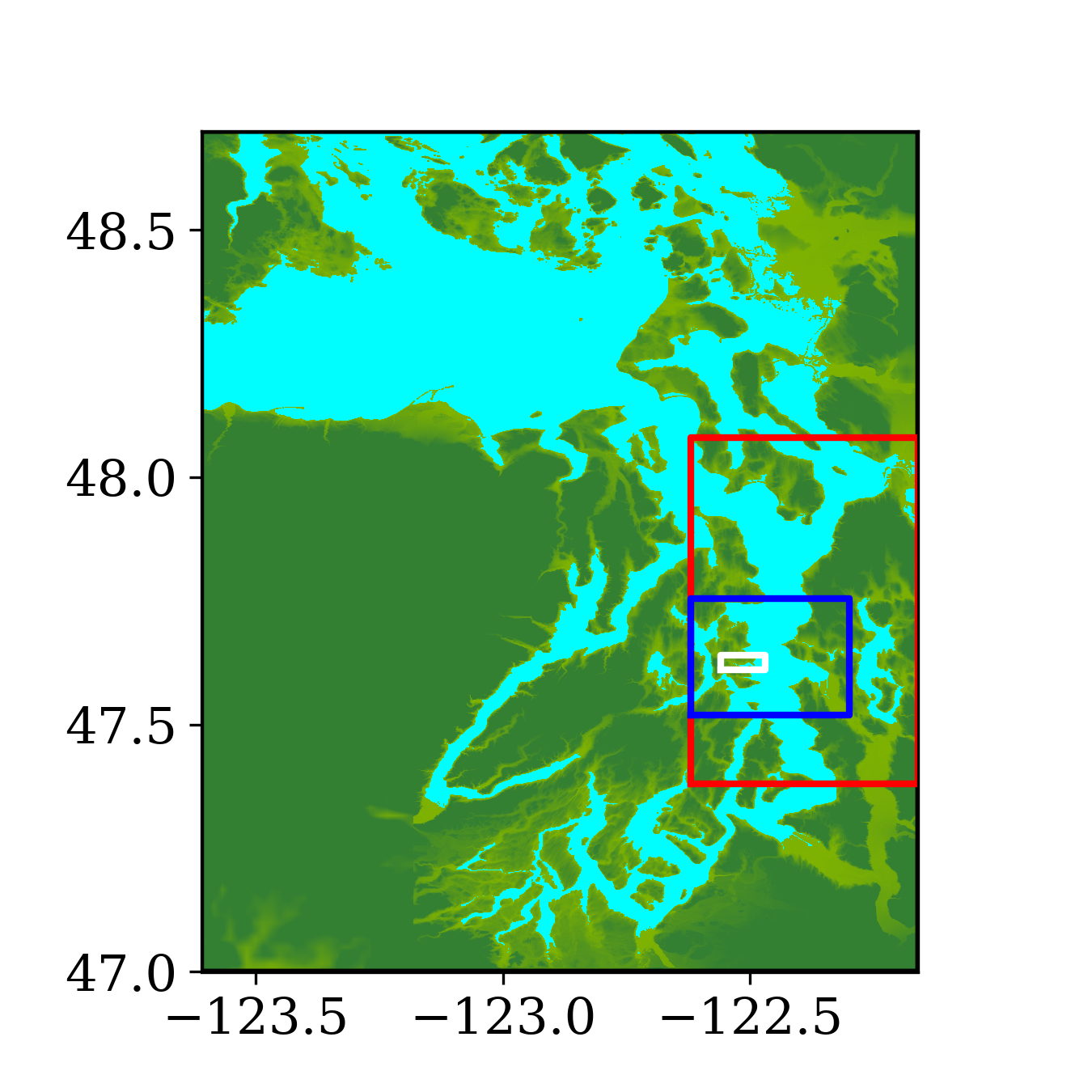} }
\hfill
\subfloat[]{\includegraphics[width=0.32\textwidth]{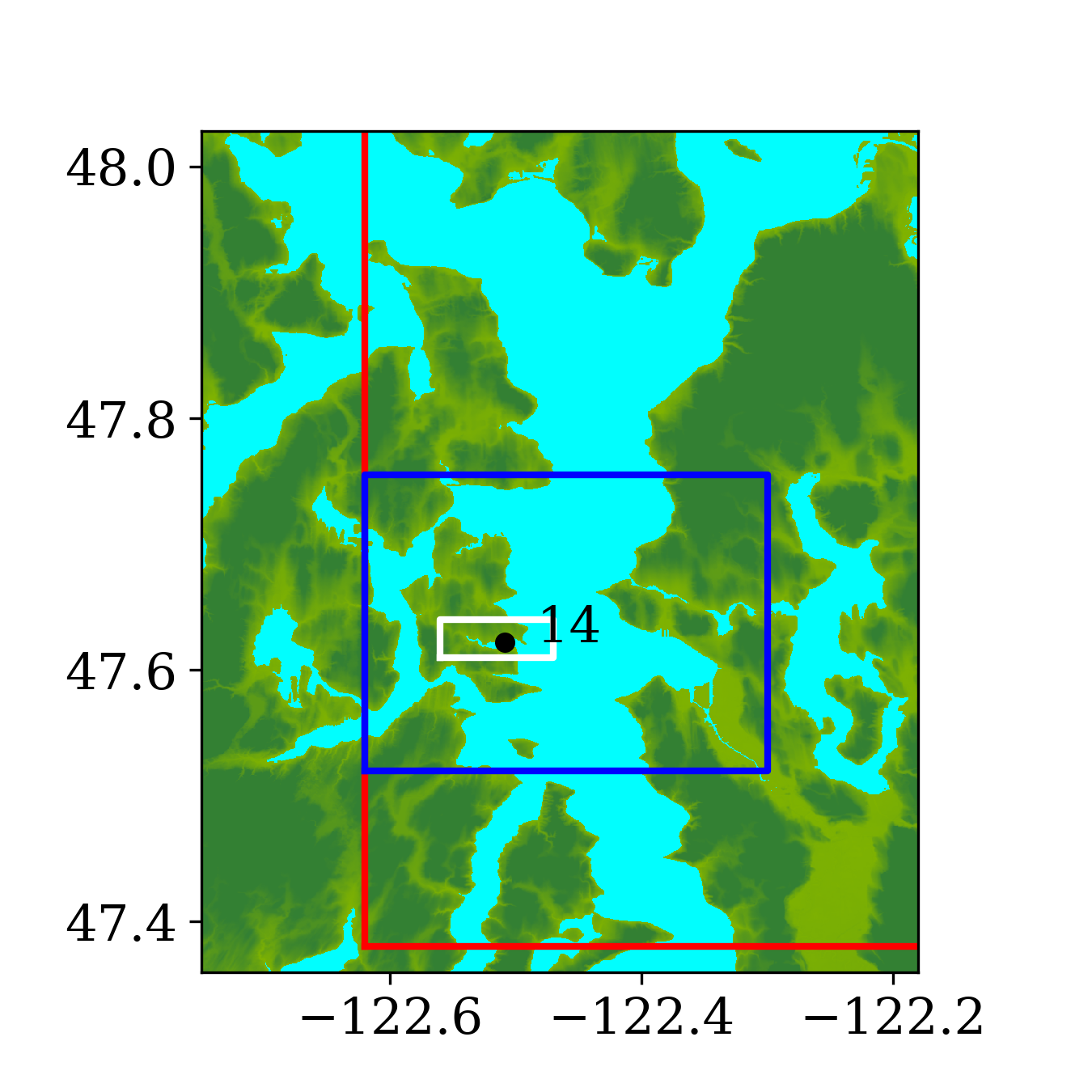} }
\hfill
\subfloat[]{\includegraphics[width=0.32\textwidth]{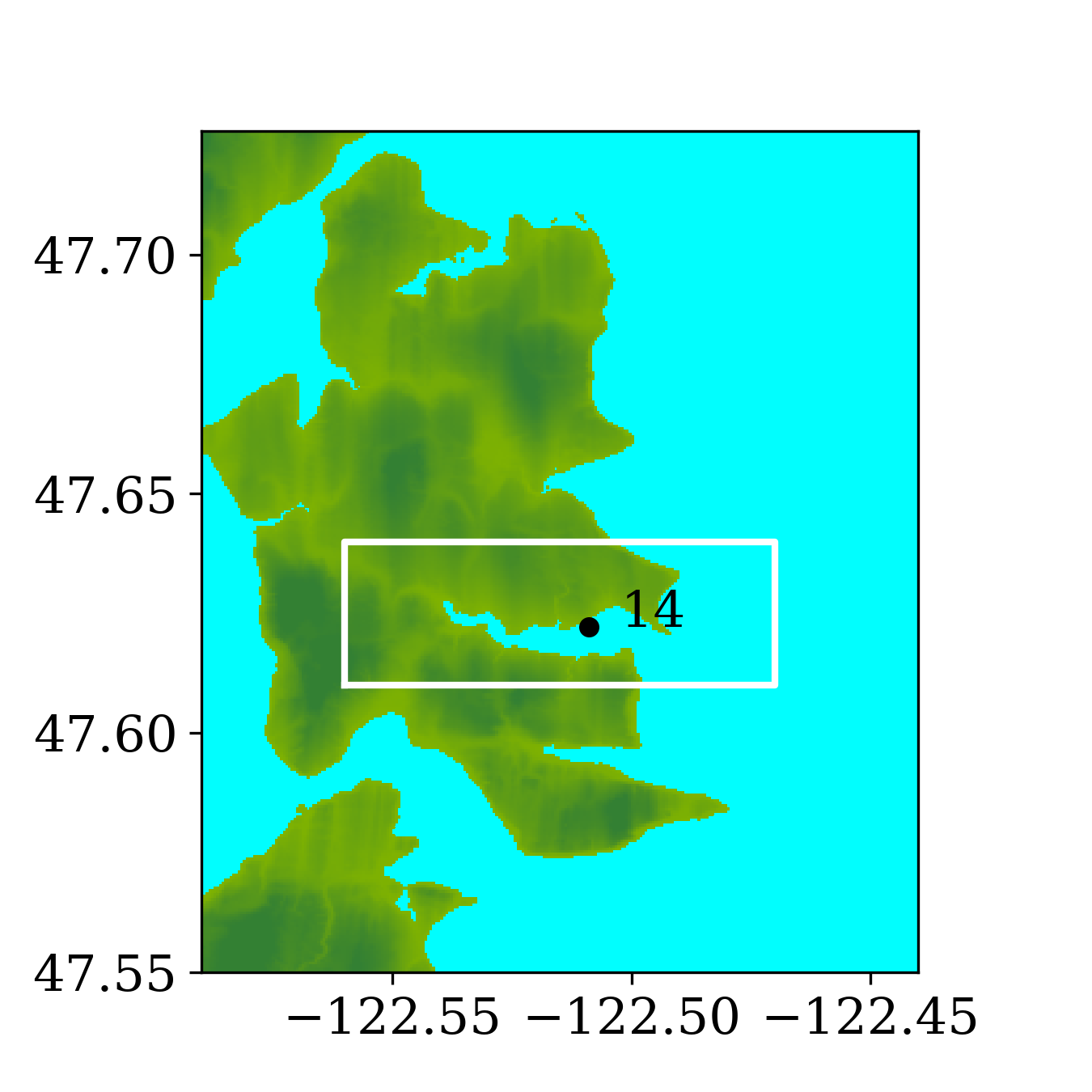} }
\caption{
    Computational domain and refinement regions for tsunami inundation triggered by the Seattle fault.
    Red rectangle shows the region where level 2 refinement is enforced.
    Blue rectangle shows the region where level 3 refinement is enforced.
    White rectangle shows the region where level 4 refinement is enforced.
     }
\label{fig:eagle_domain}
\end{figure*}

Figure \ref{fig:eagle_domain} also shows the computational domain, which is from longitude $-123.61$ to $-122.16$ and latitude 47 to 48.7.
Since this is a local tsunami in an enclosed Sound surrounded by land, the tsunami waves soon get reflected by shorelines and spread out to cover the full domain very soon after the earthquake. 
Thus, instead of using a refinement tolerance parameter, we enforce mesh refinement everywhere in the domain regardless of wave amplitude, and never regenerate new grid patches.
Four levels of refinement are used, as denoted by the rectangles in figure \ref{fig:eagle_domain} that denote regions where refinement is enforced.
Starting from the coarsest level (level 1), which has a resolution of 30 minutes, the refinement ratios are 5, 3 and 6, giving a resolution of 6 minutes on level 2, 2 minutes on level 3 and $\frac{1}{3}$ minutes on level 4.
Note that for this benchmark problem, a large proportion of the domain is dry land and the shorelines are relatively much longer and more complex.
As a result, many branches occur along the execution path of solving Riemann problems of the shallow water system since more different situations arise, e.g. a Riemann problem with one state being dry initially but becoming wet, or staying dry, depending on the flow depth and velocity in the neighboring cell.
For the GPU, if the 32 threads in a warp do not take the same execution path, each extra branch will be executed by the entire warp, introducing significant extra execution time. 
Thus the irregularity of water area in this benchmark problem is challenging for some CUDA kernels to use the GPU hardware efficiently .

\subsubsection{Simulation Results}

\begin{figure*}[t]
% from: /home2/xsqin/clawpack_cases/WA_EMD_2018/QIN/gpu/bainbridge/eagle_harbor_test/check_results/run2_30mins
\centering
\subfloat[]{\includegraphics[width=0.49\textwidth]{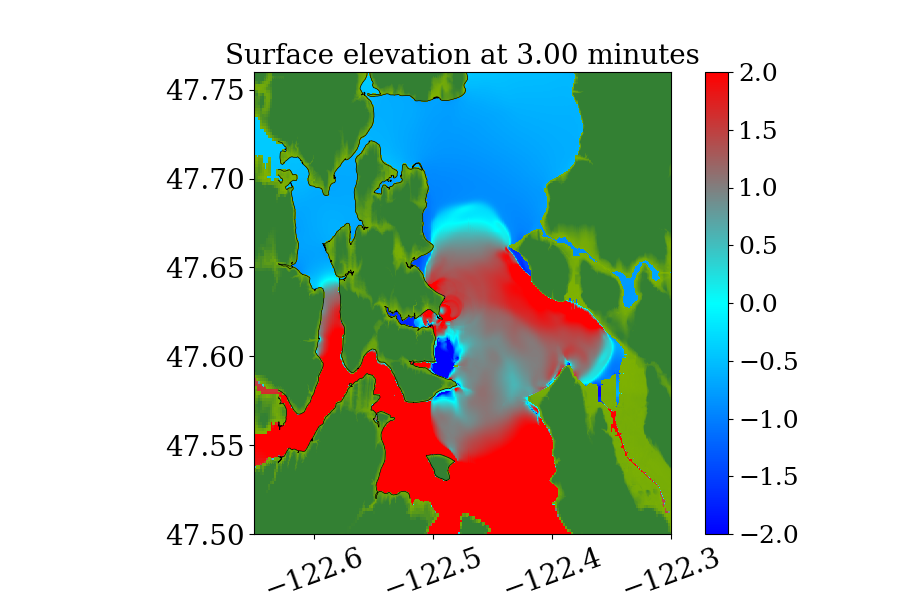} }
\hfill
\subfloat[]{\includegraphics[width=0.49\textwidth]{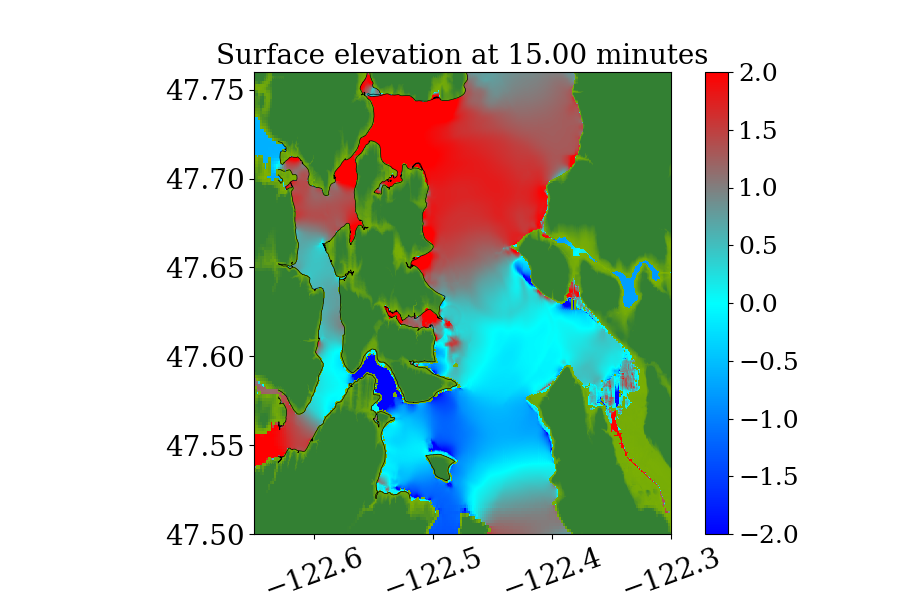} }
\caption{
    $\zeta(x,y,t)$ in Puget Sound after a tsunami triggered by Seattle fault rupture.
     }
\label{fig:simulation_eagle}
\end{figure*}

\begin{figure*}[t]
% from: /home2/xsqin/clawpack_cases/WA_EMD_2018/QIN/gpu/bainbridge/eagle_harbor_test/check_results/run2_30mins
\centering
\subfloat[]{\includegraphics[width=0.49\textwidth]{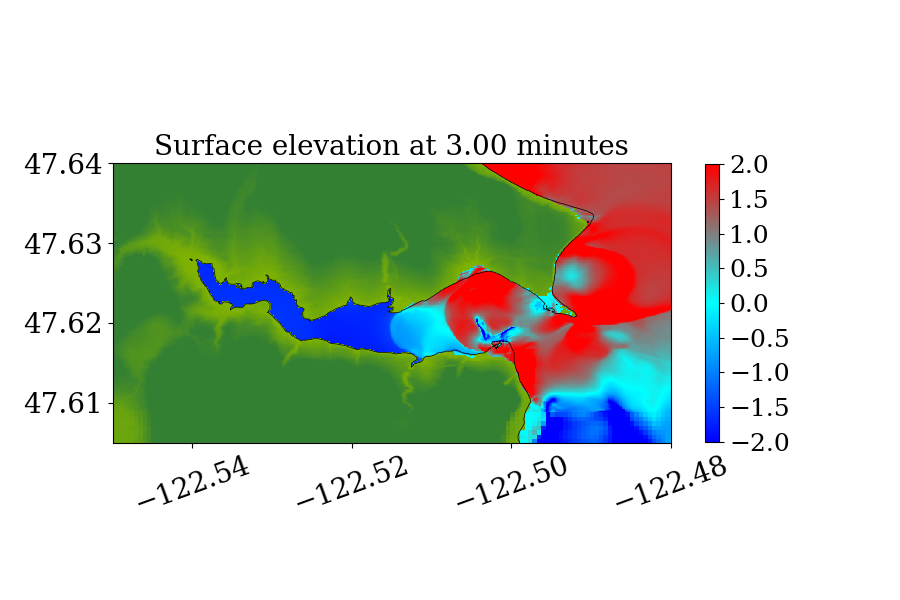} }
\hfill
\subfloat[]{\includegraphics[width=0.49\textwidth]{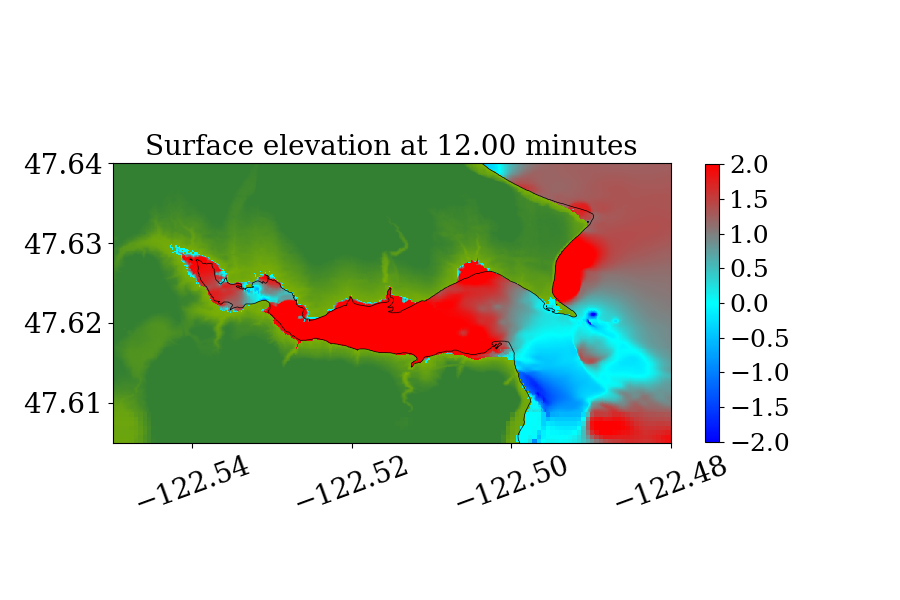} }
\caption{
    $\zeta(x,y,t)$ in Eagle Harbor of Bainbridge island after a tsunami triggered by Seattle fault rupture.
    The solid line denotes location of the shoreline at initial.
     }
\label{fig:simulation_eagle_zoom}
\end{figure*}

Figures \ref{fig:simulation_eagle} and \ref{fig:simulation_eagle_zoom} show snapshots from the simulation at several moments during the simulation in the Puget Sound and near Eagle Harbor, colored by $\zeta(x,y,t)$ defined in equation \ref{eq:zeta}.
The black solid line denotes the original shoreline before the earthquake.
At the entry of the harbor, deep inundation occurred at several places as early as only 3 minutes after the earthquake.
Even at the very end of the Eagle Harbor, the influence from the tsunami is also significant, causing more than 2-meter deep inundation in several places starting at 9 minutes after the earthquake.
One wave gauge is placed inside Eagle Harbor to record the inundation depth during the tsunami. 
Figure \ref{fig:eagle_domain} shows the location of the wave gauge.
As this is a hypothetical event for modeling an earthquake roughly 1100 year ago, there is no surface elevation observation available for comparison. 
Hence, the results from current implementation are compared with those from the MOST tsunami model \citep{titov1997implementation}.  Additional comparisons of GeoClaw and MOST results can be found in the comparison study recently performed by \citet{THA_Bainbridge2018}.  For that study the original CPU version of GeoClaw was used, with the unsplit algorithm and refluxing, but we have confirmed that very similar results are obtained with the GPU code, at least in Eagle Harbor. 

\begin{figure*}[t]
\includegraphics[width=0.9\textwidth]{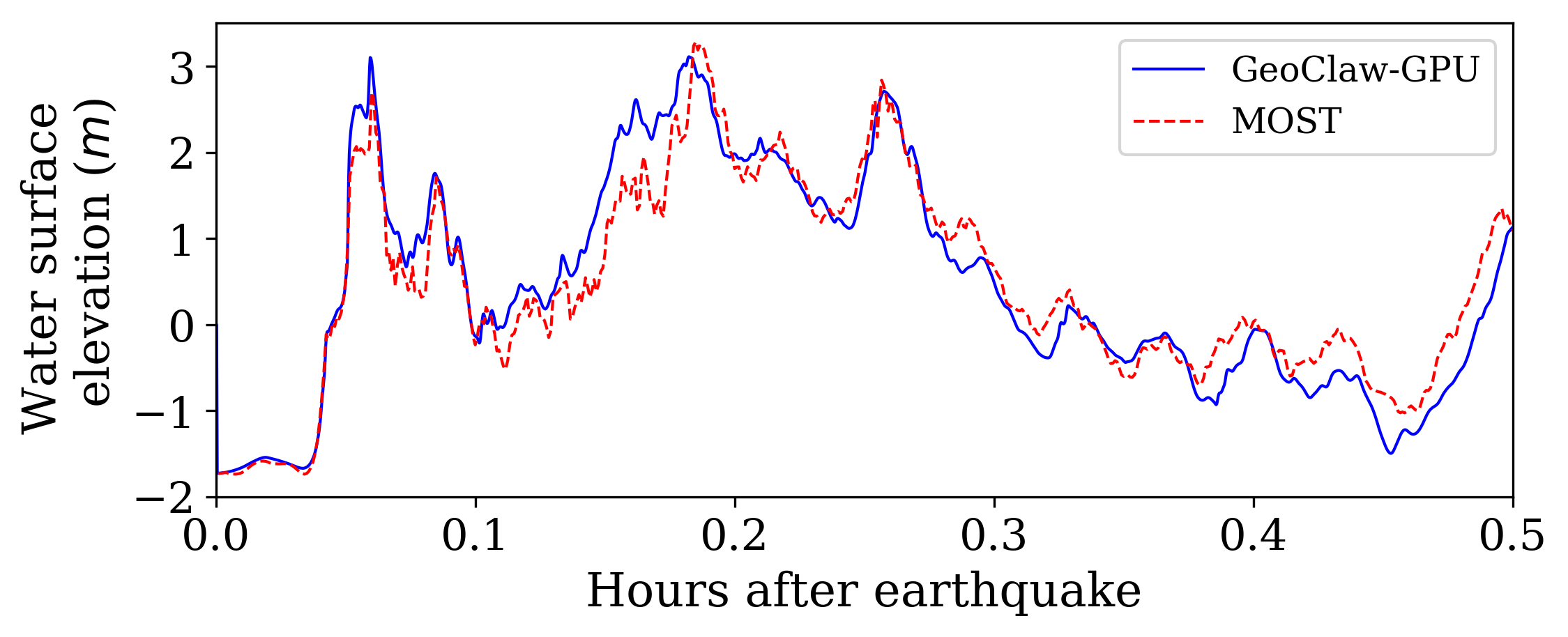}
\caption{
    Water surface elevation at a gauge (location: longitude $-122.5089$, latitude $47.6222$) inside Eagle Harbor of Bainbridge island.
}
\label{fig:wave_height_eagle}
\end{figure*}

\begin{figure}[t]
\centering
\includegraphics[width=1.0\linewidth]{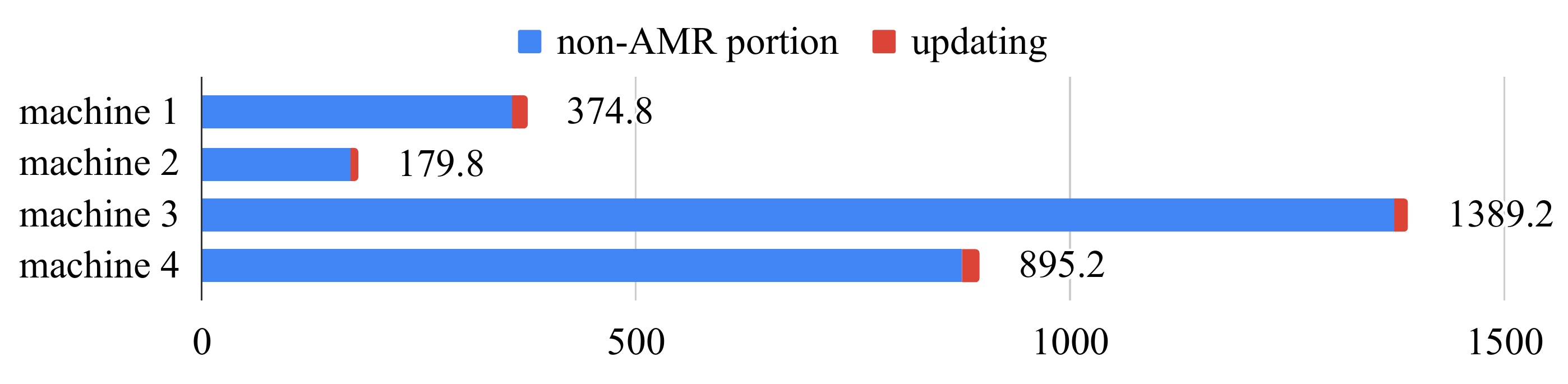}
\caption{
    Wall time (in seconds) of entire program on simulating the Seattle Fault tsunami, for original CPU implementation running on machine 3 and machine 4, and current GPU implementation running on machine 1 and machine 2. 
}
\label{fig:eagle_total_time}
\end{figure}

% \subsubsection{Relative Performance}
Figure \ref{fig:eagle_total_time} shows total running time and proportion of the two components on 4 machines (no regridding process since it is never conducted)
The total speed-ups are $3.7$ on machine 1 and $5.0$ on machine 2 for this benchmark problem.
For the original CPU implementation, the non-AMR portion takes $98\%$ and $99\%$ of the total computational time, which indicates high potential of benefiting from optimizing the performance of this portion.
Although the proportion of the non-AMR portion increases for current implementation on machine 1 and machine 2, it still takes more than $95\%$ of the total computational time, showing great potential for further improvement.

% \subsubsection{Absolute Performance}
\begin{table}[t]
\caption{
    The three metrics measured from execution of the code on machine 1 and machine 2, simulating the Seattle Fault tsunami.
}
\centering
\begin{tabular}{|l|l|l|}
\hline
   & machine 1 & machine 2 \\ \hline
$P_1$ & 60.76\%   & 57.39\%   \\ \hline
$P_2$ & 84.80\%   & 84.60\%   \\ \hline
$P_3$ & 6.87\%    & 1.77\%    \\ \hline
\end{tabular}
\label{tb:metrics_eagle}
\end{table}
Table \ref{tb:metrics_eagle} shows the three metrics for current GPU implementation running on machine 1 and machine 2 when the Seattle Fault tsunami is simulated.
Similar values are obtained for all three metrics, showing consistency and validity of the three metrics on evaluating GPU implementation with different tsunami problems.

\section{Conclusions}\label{sec:conclusions}
The shocking fatalities and infrastructure damage caused by tsunamis in the past two decades highlight the importance of developing fast and accurate tsunami models for both forecasting and hazard assessment.
This paper presents a fast and accurate GPU-based version of the GeoClaw code using patched-based AMR. 
Arbitrary levels of refinement and refinement ratios between levels are supported.
The surface elevation at DART buoys and wave gauges in benchmark problems show the ability of current tsunami model to produce accurate results in tsunami modeling.
With the GPU, the entire tsunami model runs $3.6$--$6.4$ times faster than an original CPU-based tsunami model for several benchmark problems on different machines.
As a result, the Japan 2011 Tōhoku tsunami can be fully simulated for 13 hours in under 3.5 minutes wall-clock time, using a single Nvidia TITAN X GPU. 
Three metrics for measuring the absolute performance of a GPU-based model are also proposed to evaluate current GPU implementation without comparing to others, which show the ability of current model to efficiently utilize GPU hardware resources.
Other hazards such as storm surge (e.g. \citet{MandliDawson2014}) and dam failures (e.g. \citet{George:Malpasset}) can also be modeled with GeoClaw and can similarly benefit from this GPU-enhanced version of GeoClaw.

%%%%%%%%%%%%%%%%%%%%%%%%%%%%%%%%%%%%%%%%%%%%%%%%%%%%%%%%%%%%%%%%
%
%  ACKNOWLEDGMENTS
%
% The acknowledgments must list:
%

% >>>>	A statement that indicates to the reader where the data
% 	supporting the conclusions can be obtained (for example, in the
% 	references, tables, supporting information, and other databases).
%
% 	All funding sources related to this work from all authors
%
% 	Any real or perceived financial conflicts of interests for any
%	author
%
% 	Other affiliations for any author that may be perceived as
% 	having a conflict of interest with respect to the results of this
% 	paper.
%
%
% It is also the appropriate place to thank colleagues and other contributors.
% AGU does not normally allow dedications.

\acknowledgments
The first author would like to thank Weiqun Zhang, Max Katz and Ann Almgren for many discussions with them during a summer internship at Lawrence Berkeley National Lab, supported by the AMReX project, which inspired many ideas and design strategies chosen in this work. This work was also supported in part by NSF grant EAR-1331412 and the University of Washington Department of Applied Mathematics.

%% ------------------------------------------------------------------------ %%
%% References and Citations

%%%%%%%%%%%%%%%%%%%%%%%%%%%%%%%%%%%%%%%%%%%%%%%
% BibTeX is preferred:
%
\bibliography{geo_gpu}
%
% don't specify bibliographystyle
%%%%%%%%%%%%%%%%%%%%%%%%%%%%%%%%%%%%%%%%%%%%%%%

% Please use ONLY \citet and \citep for reference citations.
% DO NOT use other cite commands (e.g., \cite, \citeyear, \nocite, \citealp, etc.).
%% Example \citet and \citep:
%  ...as shown by \citet{Boug10}, \citet{Buiz07}, \citet{Fra10},
%  \citet{Ghel00}, and \citet{Leit74}.

%  ...as shown by \citep{Boug10}, \citep{Buiz07}, \citep{Fra10},
%  \citep{Ghel00, Leit74}.

%  ...has been shown \citep [e.g.,][]{Boug10,Buiz07,Fra10}.

\end{document}